\documentclass[hyper]{JHEP} 

\usepackage{epsfig}

\newcommand\fverb{\setbox\pippobox=\hbox\bgroup\verb}

\newcommand\fverbdo{\egroup\medskip\noindent%

            \fbox{\unhbox\pippobox}\ }

\newcommand\fverbit{\egroup\item[\fbox{\unhbox\pippobox}]}

\newbox\pippobox

\title{Pure Spinor Strings in TsT Deformed
Background }

\author{
P. A. Grassi$\dag$ and J. Kluso\v{n}$\dag\dag$
 \footnote{
On leave from Masaryk University, Brno}
\footnote{\email{Josef.Kluson@roma2.infn.it}}
\\
$\dag$ DISTA, Universit\`a del Piemonte Orientale, \\
Via Bellini 25/g 15100 Alessandria, Italy, \\
 I.N.F.N. Sez. di Torino, Italy, \\
\& 
 Centro Studi e Ricerche E. Fermi, \\
Compendio Viminale, I-00184, Roma, Italy\\[1mm] 

$\dag\dag$ 
Dipartimento di Fisica \& Sezione I.N.F.N.\\
Universit\`a di Roma
``Tor Vergata'' \\
Via della Ricerca Scientifica, 1 00133  Roma   ITALY\\
}
\preprint{ \\
\hepth{}}
\abstract{We consider pure spinor strings that
propagate in the background generated by  a 
sequence of TsT transformations. We use the fact
that $U(1)$ isometry variables of TsT-transformed
background are related to the isometry variables
of the initial background in the universal way
that is independent of the details of the background.
We will argue that after redefinitions of pure spinors
and the fermionic variables we can construct pure spinor
action with manifest $U(1)$ isometry. This fact
implies that 
 the pure spinor string in TsT-transformed
background is described by pure spinor string in
the original background where world-volume modes
are subject to twisted boundary conditions. 
We will argue that these twisted boundary
conditions generally prevent to prove the
quantum conformal invariance of the
pure spinor string in $AdS_5\times S^5$ background.
We determine the conditions under which this
quantum conformal invariance can be proved.
We also determine the Lax pair for pure spinor
strings in the TsT-transformed background.}


\def\tr{\mathrm{Tr}}

\def\str{\mathrm{Str}}

\def\be{\begin{equation}}
\def\ee{\end{equation}}

\def\pb  #1{\left\{#1\right\}}

\newcommand{\hF}{\hat{F}}
\newcommand{\hG}{\hat{G}}

\newcommand{\mH}{\mathcal{H}}

\newcommand{\ow}{\overline{w}}
\newcommand{\htheta}{\hat{\theta}}

\newcommand{\halpha}{\hat{\alpha}}

\newcommand{\hgamma}{\hat{\gamma}}
\newcommand{\hlambda}{\hat{\lambda}}
\newcommand{\hw}{\hat{w}}
\newcommand{\hN}{\hat{N}}
\newcommand{\onabla}{\overline{\nabla}}

\newcommand{\tg}{\tilde{g}}

\newcommand{\com}[1]{\left[#1\right]}

\newcommand{\mL}{\mathcal{L}}

\newcommand{\uc}{\underline{c}}
\newcommand{\ud}{\underline{d}}

\newcommand{\tphi}{\tilde{\phi}}
\newcommand{\tG}{\tilde{G}}
\newcommand{\htG}{\hat{\tilde{G}}}

\newcommand{\tJ}{\tilde{J}}

\newcommand{\hg}{\hat{g}}
\newcommand{\hX}{\hat{X}}

\newcommand{\hB}{\hat{B}}
\newcommand{\tlambda}{\tilde{\lambda}}
\newcommand{\thlambda}{\tilde{\hlambda}}
\newcommand{\tw}{\tilde{w}}
\newcommand{\thw}{\tilde{\hat{w}}}
\newcommand{\hJ}{\hat{J}}
\newcommand{\bJ}{\mathbf{J}}
\newcommand{\olambda}{\overline{\lambda}}
\newcommand{\uhlambda}{\underline{\hlambda}}
\newcommand{\uhw}{\underline{\hw}}

\newcommand{\cP}{\mathcal{P}}
\newcommand{\tP}{\tilde{\cP}}
\newcommand{\hbJ}{\tilde{\bJ}}
\newcommand{\hnabla}{\tilde{\nabla}}
\newcommand{\tj}{\tilde{j}}

\begin{document}
\section{Introduction and summary}
It was noticed recently in 
\cite{Lunin:2005jy} 
 that in situation
when the initial geometry contains
a two-torus, a regular background can
be generated by using a combination of 
T-duality transformation on one angle
variable followed a shift of the second
isometry variable and finally performing
the second T-duality along the first 
isometry variable. This chain of duality
transformations that produces family
of one-parameter deformation of initial
background is known as TsT transformation.
The work \cite{Lunin:2005jy} can be generalised to
construct regular multi-parameter deformations
of gravity background if they contain
a higher dimensional torus and it is possible
to perform many chains of TsT transformations
\cite{Frolov:2005dj}.

\medskip 

Remarkable fact considering TsT transformations
is that they are very powerful for  searching of new
less supersymmetric examples of AdS/CFT correspondence.
In particular, it was used in  
\cite{Lunin:2005jy}
to obtain a
deformation of $AdS_5\times S^5$ geometry 
that is conjectured to be dual to supersymmetric
marginal deformation of $N=4$ SYM. This deformation
is called as a $\beta$ deformation. 

\medskip 

Some aspects of more general three-parameter 
deformed $AdS_5\times S^5$ and the dual
non-supersymmetric deformations of $N=4$
SYM have been studied in papers 
\cite{Berenstein:2000ux,Aharony:2002tp,Aharony:2002hx,
Dorey:2002pq,Dorey:2003pp,Niarchos:2002fc,Roiban:2003dw,
Berenstein:2004ys,Frolov:2005ty,deMelloKoch:2005vq,Mateos:2005zn,
Pal:2005nr,Freedman:2005cg,Penati:2005hp,Rossi:2005mr,Mauri:2005pa,
Kuzenko:2005gy,Ryang:2005pg,Berenstein:2005ek,Chen:2005sb,
Beisert:2005if,Frolov:2005iq,Prinsloo:2005dq,Freyhult:2005ws,Russo:2005yu,
Dymarsky:2005nc,Gauntlett:2005jb,Hernandez:2005xd,Spradlin:2005sv,
Huang:2005xh,Huang:2005mi,Bundzik:2005zg,Rashkov:2005mi}.
It is unclear, however, if the non-supersymmetric
string background is stable. For example, 
it is known that the spectrum of string theory
in the TsT-transformed flat space contains
tachyon \cite{Russo:2005yu}. However it does not
imply that string theory on the deformed
$AdS_5\times S^5$ is unstable because 
the TsT-transformed flat space is singular at space
infinity while the deformed $AdS_5\times S^5$
is regular everywhere.
As was shown in \cite{Frolov:2005dj, Alday:2005ww} 
the TsT transformation has very
nice property  
that it can be implemented on the string
sigma model leading to the simple
relations between string coordinates
of the initial and TsT dual-transformed
background.  These relations allow
to prove that the classical solutions of
string theory equations of motion in
a deformed background are in one-to one
correspondence with those in the initial
background with twisted boundary conditions
imposed on the $U(1)$ isometry fields that
parametrise the torus.
\medskip 

The analysis performed in
\cite{Frolov:2005dj} 
 was restricted to the bosonic part of type IIB 
Green-Schwarz superstring action on the
deformed $AdS_5\times S^5$. 
This work was generalised to the full
Green-Schwarz superstring action in
the remarkable paper  \cite{Alday:2005ww}.
 The problem with
superstring extension is how to define
the TsT transformation for fermion variables
since they are not neutral under T-duality
transformations
\cite{Cvetic:1999zs,Hassan:1999bv}. The  key
idea that was presented in \cite{Alday:2005ww}
 and that solves this problem is to redefine
the original fermions in such a way
that they become neutral under the isometries
of the torus.

\medskip 

The goal of this paper is to see that the
same analysis can be performed in
case of pure spinor string proposed  
by Berkovits
\cite{Berkovits:1996bf,Berkovits:2000fe,
Berkovits:2000ph,Berkovits:2000nn, Berkovits:2001us}
 \footnote{For
review of pure spinor formalism in 
superstring theory, see
\cite{Berkovits:2002zk,Grassi:2005av,
Grassi:2003cm,Grassi:2002sr,
Nekrasov:2005wg}.}. 
In a recent paper \cite{Berkovits:2004xu},
quantum consistency was argued by means of algebraic
renormalization arguments.
The one-loop conformal invariance of pure spinor string
was also demonstrated  in 
\cite{Vallilo:2002mh}
\footnote{Check  of the one-loop conformal
invariance of pure spinor string in
general background was performed in
\cite{Chandia:2003hn,Bedoya:2006ic}.}.
 Vertex operators for massless
excitations have been proposed some time ago
\cite{Berkovits:2000yr} and checked to be classically BRST
invariant \cite{Kluson:2006wq}.  Algebra of currents was
also classically calculated in 
\cite{Bianchi:2006im}
 and the
first attempt to calculate their operator product 
expansion was performed in \cite{Puletti:2006vb}.

\medskip 
 
All these results, especially proof of the quantum
consistency of the pure spinor string in $AdS_5\times S^5$
suggest that pure spinor string could be the
correct way to  study the string theory on the 
$\gamma$-deformed background. The goal of this
paper is to demonstrate this fact. Let us outline
its  content. 

\medskip 

We will show that we can 
formulate the pure spinor
string in the deformed background using the TsT
transformations from the original $AdS_5\times S^5$
background. As in the case of
GS superstring \cite{Alday:2005ww}
we redefine
both fermions and pure spinors variables in order
that they become neutral under 
isometry transformations. 
Then we argue that the pure spinor string in 
$\gamma$-deformed $AdS_5\times S^5$ 
background is equivalent
to the pure spinor string in the original $AdS_5\times
S^5$ background where  the 
 world-sheet fields
obey twisted boundary conditions. 
We will also argue that 
 the existence of these twisted 
boundary conditions is crucial for the proof
of the quantum consistency of the pure spinor string
in $\gamma$ deformed background. 
More precisely, the proof
of the quantum consistency of pure spinor
string in $AdS_5\times S^5$ presented
in  \cite{Berkovits:2004xu} was based on the
explicit gauge invariance of the pure spinor
string in $AdS_5\times S^5$ background. On the other
hand the twisted boundary conditions for
world-sheet fields are naturally related
to the particular  coset representative  that
however breaks the explicit gauge invariance
of the theory. Then we will show that  
 in  order to restore this 
gauge invariance we have to restrict to the
case when the world-sheet fields 
obey periodic boundary conditions. 

\medskip 

Let us be more explicit.  We will
see  that the configuration of the
pure spinor string in 
 $AdS_5\times S^5$ background
is labelled in general by $3$ conserved
angular momenta $(J_1,J_2,J_3)$. These angular
momenta depend on the deformation parameters
$\gamma_i$ through 
\begin{equation}
\nu_i\equiv \epsilon_{ijk}\gamma_j J_k \ . 
\end{equation}
These combinations are the twists that
appear in the relations between
the angle variables of $S^5$ and the $
\gamma_i$-deformed sphere. We will
argue that for $\nu_i$ equal to integer
the currents of the pure spinor string
in the $AdS_5\times S^5$ background
obey the periodic boundary conditions
and hence the gauge invariance of the
theory can be restored. This result implies
that the states of the pure spinor
string in the $\gamma$ deformed
background that obey the condition
$\nu_i$ is integer correspond to the
string theory in  $AdS_5\times S^5$
background that pose the gauge invariance
of the coset and that, according to the
arguments by  N. Berkovits given  
\cite{Berkovits:2004xu} has
exact conformal invariance. This
result confirms the analysis performed in 
\cite{Frolov:2005iq}. We hope that
the arguments given in this paper suggests
that states with $\nu_i$ equal to integer
in $\gamma_i$-deformed background have
exact conformal field theory description. 
\medskip 

Let us outline the structure of the paper. In next
section  (\ref{second}) we review  how the
TsT transformation is defined in the context
of the non-linear sigma model. In section
(\ref{third}) we  introduce the action for
pure spinor string in $AdS_5\times S^5$ background.
Then we  determine its form  using
the explicit parametrisation of the coset introduced in 
\cite{Alday:2005gi}. 
 In section (\ref{fourth})
we study the equations of motions for pure spinor
string in the coset representation. We 
 prove the conservation of the BRST currents.
In section   (\ref{fifth}) we perform the  redefinition
of the fermions and pure spinor variables 
following \cite{Alday:2005ww}.

Then in section (\ref{sixth}) we apply 
TsT transformation to the five sphere and
we find the relation between the pure spinor
string action in $\gamma$-deformed action and
in the original $AdS_5\times S^5$ action.   
Finally, in section (\ref{seventh}) we  argue
for an existence of the Lax connection for pure
spinor string in the $\gamma$ deformed background
again following the approach given in
\cite{Alday:2005ww}.
\section{Review of the $
\gamma$-deformed action} \label{second}
 We start
with the sigma model action that
describes the propagation of closed
string on the background with several
$U(1)$ isometries
\begin{eqnarray}\label{sigmaphi}
S&=&-\frac{\sqrt{\lambda}}{4\pi} \int d
\tau d\sigma \sqrt{-h}[h^{\mu\nu}
\partial_\mu \phi^i\partial_\nu
\phi^j G_{ij}^0-\epsilon^{\mu\nu}
\partial_\mu \phi^i\partial_\nu
\phi^j B_{ij}^0+\nonumber \\
&+& 2\partial_\mu
\phi^i(h^{\mu\nu}U_{\nu,i}^0-
\epsilon^{\mu\nu}V_{\nu,i}^0)+
\mathcal{L}^0_{rest}] \ . \nonumber \\
\end{eqnarray}
As usual we have introduced the
effective string tension
$\frac{\sqrt{\lambda}}{2\pi}$ that is
identified with the 't Hooft coupling
in the AdS/CFT correspondence, $h_{\mu\nu}$
is worldsheet metric with Minkowski
signature  that in conformal gauge is
$h^{\mu\nu}=(-1,1)$ and
$\epsilon^{\mu\nu}=\frac{e^{\mu\nu}}{
\sqrt{-h}} \ ,
e^{01}=-e^{10}=e^{\tau\sigma}=1$. Next
we assume that the action is invariant
under the $U(1)$ isometry
transformations that are geometrically
realised as shifts of the angle
variables $\phi_i \ , i=1,2,\dots,d$.
In other words the string background
contains the $d$-dimensional torus
$T^d$. The action (\ref{sigmaphi})
explicitly shows the dependence
on $\phi^i$ and their coupling to the
background fields $G_{ij}^0 \ ,
B_{ij}^0$ and
$U_{\nu,i}^0,V_{\nu,i}^0$. These
background fields are independent on
$\phi^i$ but can depend on other
bosonic and fermionic string
coordinates which are neutral under the
$U(1)$ isometry transformations.
Finally $\mathcal{L}^0_{rest}$ denotes
the part of the Lagrangian that depends
on other fields of the theory. 

As previous  discussion suggests the
 action (\ref{sigmaphi}) is
invariant under the constant shift of
$\phi^i$
\begin{equation}
\phi'^i(\tau,\sigma)=\phi^i(\tau,\sigma)+
\epsilon^i \ .
\end{equation}
The corresponding  Noether currents have
the form
\begin{equation}\label{current}
J^\mu_i=-\frac{\sqrt{\lambda}}{2\pi}
\sqrt{-h}(h^{\mu\nu}
\partial_\nu \phi^j G_{ji}^{0}
-\epsilon^{\mu\nu}\partial_\nu
\phi^jB_{ij}^0+h^{\mu\nu}U_{\nu,i}^0-
\epsilon^{\mu\nu}V^0_{\nu,i}) \  
\end{equation}
and obeys the equation 
\begin{equation}\label{paju}
\partial_\mu J^\mu_i=0 \  
\end{equation}
as a consequence of the equations of 
motion. 

Now we are ready to study TsT duality of the
angle variables. Let us consider
 two-torus that is generated by
$\phi_1$ and $\phi_2$. 
The TsT transformation consists 
T-dualizing the variable $\phi_1$
with the further shift
$\phi_2\rightarrow
\phi_2+\tilde{\gamma}\phi_1$ and
dualizing $\phi_1$ back. The
TST transformation can be symbolically
expressed as 
\begin{equation}
(\phi_1,\phi_2)
\stackrel{\rm TsT}{\rightarrow}
   (\tphi_1,
\tphi_2) \ . 
\end{equation}
In order to find the TsT transformation
of the non-linear sigma model action
we proceed  following
the classical works    
\cite{Buscher:1987sk,Buscher:1987qj}.

Let us start with the T-duality on a circle
parametrised by $\phi_1$. As the next
step we   gauge the shift  symmetry 
$\phi'^1=\phi^1+\epsilon^1$
so that $\epsilon^1$ is now
function of  $\tau,\sigma$.
If we require that the action is
invariant under the non-constant
transformation we have to 
introduce  the appropriate
gauge field $A_\mu$ in such
a way that
\begin{equation}
\partial_\mu \phi^1
\rightarrow (\partial_\mu \phi^1
+A_\mu)\equiv D_\mu \phi^1 \ . 
\end{equation}
At the same time we add to the
action the term $\tphi^1 
\epsilon^{\mu\nu}F_{\mu\nu}
$ in order to assure that the
gauge field has trivial dynamics.
Then we obtain the gauge invariant
action  
\begin{eqnarray}\label{sigmaphig}
S&=&-\frac{\sqrt{\lambda}}{4\pi} \int d
\tau d\sigma \sqrt{-h}[h^{\mu\nu}
D_\mu \phi^1D_\nu
\phi^1 G_{11}^0
+2h^{\mu\nu}D_\mu \phi^1\partial_\nu
\phi^a G_{1a}^0
+h^{\mu\nu}\partial_\mu \phi^a\partial_\nu
\phi^b G_{ab}^0
-\nonumber \\
&-&\epsilon^{\mu\nu}
\partial_\mu \phi^a\partial_\nu
\phi^b B_{ab}^0
-2\epsilon^{\mu\nu}
D_\mu \phi^1\partial_\nu
\phi^b B_{1b}^0
+\nonumber \\
&+&2D_\mu
\phi^1(h^{\mu\nu}U_{\nu,1}^0-
\epsilon^{\mu\nu}V_{\nu,1}^0)+
2\partial_\mu
\phi^a(h^{\mu\nu}U_{\nu,a}^0-
\epsilon^{\mu\nu}V_{\nu,a}^0)+
\tilde{\phi}^1\epsilon^{\mu\nu}
F_{\mu\nu}+
\mathcal{L}^0_{rest}] \ , \nonumber \\
\end{eqnarray}
where  $a,b=2,\dots, d$. 
Now thanks to the gauge invariance
we can fix the gauge $\phi^1=0$ so
that the action above takes the form
\begin{eqnarray}\label{sigmaphigf}
S&=&-\frac{\sqrt{\lambda}}{4\pi} \int d
\tau d\sigma \sqrt{-h}[h^{\mu\nu}
A_\mu A_\nu
 G_{11}^0
+2h^{\mu\nu}A_\mu \partial_\nu
\phi^a G_{1a}^0
+h^{\mu\nu}\partial_\mu \phi^a\partial_\nu
\phi^b G_{ab}^0
-\nonumber \\
&-&\epsilon^{\mu\nu}
\partial_\mu \phi^a\partial_\nu
\phi^b B_{ab}^0
-2\epsilon^{\mu\nu}
A_\mu \partial_\nu
\phi^b B_{1b}^0
+\nonumber \\
&+& 2A_\mu
(h^{\mu\nu}U_{\nu,1}^0-
\epsilon^{\mu\nu}V_{\nu,1}^0)+
2\partial_\mu
\phi^a(h^{\mu\nu}U_{\nu,a}^0-
\epsilon^{\mu\nu}V_{\nu,a}^0)+
\tilde{\phi}^1\epsilon^{\mu\nu}
F_{\mu\nu}+
\mathcal{L}^0_{rest}] \ . \nonumber \\
\end{eqnarray}
If we now integrate $\tilde{\phi}^1$
we obtain that $F_{\mu\nu}=0$
and hence $A_\mu=\partial_\mu
\theta$. Inserting back to the action
(\ref{sigmaphigf}) we obtain
the original action (\ref{sigmaphi})
after identification $\theta=\phi^1$. 
On the other hand if we integrate out
$A_\mu$  we obtain
\begin{eqnarray}
2h^{\mu\nu}A_\nu G^0_{11}+
2h^{\mu\nu}\partial_\nu \phi^a G_{1a}^0
-2\epsilon^{\mu\nu}\partial_\nu
\phi^a B_{1a}^0+2(h^{\mu\nu}
U^0_{\nu,1}-\epsilon^{\mu\nu}V_{\nu,1})
-2\partial_\nu [
\epsilon^{\nu\mu}\tilde{\phi}^1]=0
\nonumber \\
\end{eqnarray}
that implies
\begin{equation}\label{Aphi}
A_\mu=\frac{1}{G^0_{11}}
(-\partial_\mu \phi^a G_{1a}^0
+h_{\mu\nu}\epsilon^{\nu\rho}
\partial_\rho\phi^a B_{1a}^0
-(U^0_{\mu,1}-h_{\mu\nu}\epsilon^{\nu\rho}
V_{\rho,1}^0)-h_{\mu\nu}
\epsilon^{\nu\rho}
\partial_\rho\tilde{\phi} ) \ . 
\end{equation}
Since we have argued that $A_\mu$
can be related to the original
coordinate $\phi^1$ as $A_\mu=\partial_\mu\phi^1$
the relation (\ref{Aphi}) implies following
relation between $\phi^1$ and $\tphi^1$
\begin{eqnarray}\label{relphitphih}
\epsilon^{\nu\rho}\partial_\rho
\tphi^1 &=&-h^{\nu\rho}
G_{11}^0\partial_\rho\phi^1
-h^{\nu\rho}
\partial_\rho \phi^a G_{1a}^0
+\epsilon^{\nu\rho}
\partial_\rho\phi^a B_{1a}^0
-h^{\nu\rho}
U^0_{\rho,1}+\epsilon^{\nu\rho} V_{\rho,1}^0 \ , \nonumber
\\
\tphi^a&=&\phi^a \ . 
\end{eqnarray}
Now plugging the result
(\ref{Aphi})
 into the action
above we obtain the action 
equivalent to (\ref{sigmaphi})
\begin{eqnarray}\label{sigmaphid}
S&=&-\frac{\sqrt{\lambda}}{4\pi} \int d
\tau d\sigma \sqrt{-h}[h^{\mu\nu}
\partial_\mu \tphi^i\partial_\nu
\tphi^j \tilde{G}_{ij}-\epsilon^{\mu\nu}
\partial_\mu \tphi^i\partial_\nu
\tphi^j \tilde{B}_{ij}+\nonumber \\
&+&2\partial_\mu
\phi^i(h^{\mu\nu}\tilde{U}_{\nu,i}-
\epsilon^{\mu\nu}\tilde{V}_{\nu,i})+
\tilde{\mathcal{L}}_{rest}] \ , \nonumber \\
\end{eqnarray}
where now
\begin{eqnarray}
\tilde{G}_{11}&=&\frac{1}{G^0_{11}} \ , \quad
\tilde{G}_{ab}=G_{ab}^0-
\frac{G_{a1}^0G_{1b}^0-B_{1a}^0
B_{1b}^0}{G_{11}^0} \ ,  \quad
\tilde{G}_{1a}=\frac{B_{1a}^0}{G_{11}^0} \ ,
\nonumber \\
\tilde{B}_{ab}&=&B_{ab}^0-\frac{G_{1a}^0B_{1b}^0
-B_{1a}^0G^0_{1b}}{G_{11}^0} \ , \quad
\tilde{B}_{1a}=\frac{G_{1a}^0}{G_{11}^0} \ ,
\quad  
\tilde{B}_{a1}=-\frac{G_{1a}^0}{G_{11}^0} \ , 
\nonumber \\
\tilde{U}_{\mu,1}&=&\frac{V_{\mu,1}^0}{G^0_{11}} \ , 
\quad 
\tilde{V}_{\mu,1}=\frac{U_{\mu,1}^0}{G^0_{11}} \ , 
\nonumber \\
\tilde{U}_{\mu,a}&=&U_{\mu,a}^0-
\frac{G^0_{1,a}U^0_{\nu,1}-B_{1a}^0V^0_{\mu,1}
}{G_{11}^0} \ , \nonumber \\
\tilde{V}_{\mu,a}&=&
V^0_{\mu,a}-\frac{G_{1a}^0V_{\mu,1}^0-
B^0_{1a}U^0_{\mu,1}}{G_{11}^0} \ , 
\nonumber \\
\tilde{\mathcal{L}}_{rest}&=&
\mathcal{L}^0_{rest}-
h^{\mu\nu}\frac{U^0_{\mu,1}U^0_{\nu,1}-
V^0_{\mu,1}V^0_{\nu,1}}{G_{11}^0}+
\epsilon^{\mu\nu}
\frac{U^0_{\mu,1}V^0_{\nu,1}-V^0_{\mu,1}U^0_{\nu,1}}
{G_{11}^0} \ . \nonumber \\
\end{eqnarray}
Clearly the action 
(\ref{sigmaphid}) has the same
number of symmetries as the
original one. 

The next step in the definition
of the TsT transformation
is the  the shift
of the variables $\tphi^a$
that is defined as
\begin{eqnarray}\label{shiftT}
\tphi^2 &=&
\tphi^2_s+\hgamma\tphi^1_s, \quad
\tphi^1_s=\tphi^1 \ ,  \nonumber \\
\tphi^a_s&=&\tphi^a \ , 
\quad a=3,\dots, d \ . \nonumber \\
\end{eqnarray}  
If we now insert 
(\ref{shiftT})
into the action (\ref{sigmaphid}) we
get 
\begin{eqnarray}\label{sigmaphidss}
S&=&-\frac{\sqrt{\lambda}}{4\pi} \int d
\tau d\sigma \sqrt{-h}[h^{\mu\nu}
\partial_\mu \tphi^i_s\partial_\nu
\tphi^j_s \tilde{G}^s_{ij}-\epsilon^{\mu\nu}
\partial_\mu \tphi^i_s\partial_\nu
\tphi^j_s \tilde{B}^s_{ij}+\nonumber \\
&+&2\partial_\mu
\tphi^i_s(h^{\mu\nu}\tilde{U}^s_{\nu,i}-
\epsilon^{\mu\nu}\tilde{V}^s_{\nu,i})+
\tilde{\mathcal{L}}_{rest}] \ , \nonumber \\
\end{eqnarray}
where the forms of the background
fields $\tilde{G}_{ij}^s,\ \tilde{B}^s_{ij}, \ 
\tilde{U}^s_{\nu,i}$ and 
$\tilde{V}^s_{\nu,i}$ can be easily
determined from the action 
(\ref{sigmaphid}) and the shift
transformation (\ref{shiftT}).
Finally   
 we perform the last
 T-duality transformation along the
direction labelled with $\tphi^1_s$.
 After this transformation
we get the action in the final form
\begin{eqnarray}\label{sigmaphids}
S&=&-\frac{\sqrt{\lambda}}{4\pi} \int d
\tau d\sigma \sqrt{-h}[h^{\mu\nu}
\partial_\mu \phi^i_F\partial_\nu
\phi^j_F G_{ij}-\epsilon^{\mu\nu}
\partial_\mu \phi^i_F\partial_\nu
\phi^j_F B_{ij}+\nonumber \\
&+& 2\partial_\mu
\phi^i_F(h^{\mu\nu}U_{\nu,i}-
\epsilon^{\mu\nu}V_{\nu,i})+
\mathcal{L}_{rest}] \ , \nonumber \\
\end{eqnarray}
where now 
\begin{eqnarray}
G_{ij}&=&
\frac{G_{ij}^0}{D} \ , \quad
G_{ia}=G_{ai}
=\frac{G_{ia}^0}{D}
+\hgamma\frac{B_{2a}^0 G_{1i}^0-B_{1a}^0
G_{2i}^0+B_{12}^0G_{ia}^0}{D} \ ,
\nonumber \\
\nonumber \\
G_{ab}&=&G_{ab}^0
+\frac{\hgamma+\hgamma^2B_{12}^0}{D}
2(B_{23}^0G_{13}^0-B_{13}^0G_{23}^0)+
\nonumber \\
&+&\frac{\hgamma^2}{D}\left(
G_{11}^0(B_{2a}^0B_{2b}^0-G_{2a}^0G_{2b}^0
+G_{22}^0(B_{1a}^0B_{1b}^0-G_{1a}^0G_{1b}^0)
+2G_{12}^0(G_{2a}^0G_{1b}^0-B_{1a}^0
B_{2b}^0)\right)
\nonumber \\
\end{eqnarray}
and
\begin{eqnarray}
B_{12}&=&-B_{21}
=\frac{B_{12}^0}{D}+\frac{\hgamma}{D}
(G_{11}^0G_{22}^0-(G_{12}^0)^2+
(B_{12}^0)^2) \ ,
\nonumber \\
B_{ia}&=&-B_{ai}=\frac{B_{ia}^0}{D}+
\frac{\hgamma}{D}(G_{2a}^0G_{i1}^0
-G_{i2}^0G_{1a}^0+B_{12}^0
B_{ia}^0) \ , 
\nonumber \\
U_{\mu,i}
&=&\frac{U^0_{\mu,i}}{D}+
\frac{\hgamma}{D}
(G_{11}^0 V_{\mu,2}^0-G_{2i}^0
V_{\mu,1}^0+B^0_{12}U^0_{\mu,i}) \ , 
\nonumber \\
V_{\mu,i}&=&
\frac{V_{\mu,i}^0}{D}
+\frac{\hgamma}{D}
(B_{12}^0V_{\mu,i}^0+
G_{1i}^0U_{\mu,2}^0
-G_{2i}^0U_{\mu,1}^0) \ , 
\nonumber \\
U_{\mu,a}&=&
U_{\mu,a}^0+\frac{(\hgamma+
\hgamma^2 B_{12}^0)}
{D}(\epsilon^{ij}G_{ia}^0V_{\mu, j}^0
-\epsilon^{ij}B_{ia}^0U_{\mu,j}^0)+
\nonumber \\
&+&\frac{\hgamma^2}{D}
(\epsilon^{ij}U_{\mu,i}^0(G_{2a}^0
G_{1j}^0-G_{1a}^0G_{2j}^0)+\epsilon^{ij}
V_{\mu,i}^0(-B_{2a}^0G_{1j}^0
+B_{1a}^0G_{2j}^0)) \ , 
\nonumber \\
V_{\mu,a}&=&
V_{\mu,a}^0+\frac{(\hgamma+\hgamma^2B_{12}^0)}
{D}(\epsilon^{ij}G_{i3}^0U_{\mu,j}^0
-\epsilon^{ij}B_{ia}^0V_{\mu, j}^0)
+\nonumber \\
&+&\frac{\hgamma^2}{D}
(\epsilon^{ij}V_{\mu,i}^0(G_{2a}^0
G_{1j}^0-G_{1a}^0G_{2j}^0)+\epsilon^{ij}
U_{\mu,i}^0(-B_{2a}^0G_{1j}^0+
B_{1a}^0G_{2j}^0)) \ , 
\nonumber \\ 
\mathcal{L}_{rest}&=&
\mL_{rest}^0+\frac{(\hgamma+
\hgamma^2 B_{12}^0)}{D}
(2\epsilon^{ij}(V_{0,i}^0V_{1,j}^0
-U_{0,i}^0U_{1,j}^0+h^{\mu\nu}U_{\mu,i}^0
V_{\nu,j}^0))+
\nonumber \\
&+&\frac{\hgamma^2}{D}
(G_{ij}^0\epsilon^{i\overline{i}}
\epsilon^{j\overline{j}}h^{\mu\nu}
(V_{\mu,\overline{i}}^0V_{\nu,\overline{j}}^0
-U_{\mu,\overline{i}}^0U_{\nu,\overline{j}}^0)
+G_{ij}^0\epsilon^{i\overline{i}}\epsilon^{j\overline{j}}
\epsilon^{\mu\nu}U_{\mu\overline{i}}^0
V_{\nu\overline{j}}^0) \ ,  \nonumber \\
\end{eqnarray}
where $i,j=1,2$ 
define the directions of a two-torus
and the index $a$ runs over $3,\dots,d$. 
The element $D$ is given by
\begin{equation}
D=1+2\hgamma G_{12}^0+
\hgamma^2(G_{11}^0G_{22}^0-(G_{12}^0)^2
+(B_{12}^0)^2) \ , \quad 
\hgamma=\sqrt{\lambda}\gamma \ . 
\end{equation}
Repeating the arguments given
below the first T-duality transformation
we can find the relation between
between $\phi_F$ and
$\phi$ in the form 
\begin{eqnarray}\label{relphiF}
\partial_\mu \phi^1_F&=&
\partial_1\phi^1-\hgamma \epsilon_{\mu\nu}
h^{\nu\rho}\partial_\rho \phi^i G_{i2}+
\hgamma \partial_\mu \phi^i B_{i2}-
\hgamma \epsilon_{\mu\nu}h^{\nu\rho}
U_{\rho 2}-\hgamma V_{\mu 2} \ , 
\nonumber \\
\partial_\mu\phi_F^2&=&
\partial_\mu\phi^2+\hgamma\epsilon_{\mu\nu}
h^{\nu\rho}\partial_\rho \phi^i G_{i 1}
-\hgamma \partial_\mu \phi^i B_{i1}+
\hgamma\epsilon_{\mu\nu}h^{\nu\rho}
U_{\rho 1}+\hgamma V_{\mu 1} \ , \quad 
i,j=1,\dots d  \ , 
\nonumber \\
\partial_\mu\phi^a_F&=&
\partial_\mu \phi^a \ , a=3,\dots, d \ .  
\nonumber \\
\end{eqnarray}
In what follows we rename $\phi^F$ as
$\tphi$ in order to have contact
with \cite{Alday:2005ww}. 
Clearly the action (\ref{sigmaphids})
has the  same number of symmetries as
related to the the constant
shifts of the variables $\tphi$. The
conserved Noether currents have the
form
\begin{equation}\label{currentT}
\tJ^\mu_i=-\frac{\sqrt{\lambda}}{2\pi}
\sqrt{-h}(h^{\mu\nu}
\partial_\nu \tphi^j G_{ji}
-\epsilon^{\mu\nu}\partial_\nu
\tphi^jB_{ij}+h^{\mu\nu}U_{\nu,i}-
\epsilon^{\mu\nu}V_{\nu,i}) \  .  
\end{equation}
It is important to stress that 
following relations holds 
\cite{Frolov:2005dj,Alday:2005ww}
\begin{equation}\label{eqJ}
\tJ^{\mu}_i(\tphi)=J^\mu_i(\phi) \ . 
\end{equation}
Now using (\ref{relphiF}),
(\ref{eqJ}) together
with (\ref{current}) and 
(\ref{currentT}) we obtain
\begin{eqnarray}\label{relphitphi}
\partial_1 \tphi^1-\partial_1 \phi^1&=&
-\gamma J^\tau_2 \ ,
\nonumber \\ 
\partial_1 \tphi^2-\partial_1\phi^1&=&
\gamma J^\tau_ 1 \ , 
\nonumber \\
\partial_1\tphi^i-\partial_1\phi^i &=&0 \ , \quad  
i >2 \ . \nonumber \\
\end{eqnarray}
Since we consider the closed string
on the $\gamma$-deformed background 
the angle variables $\tphi^i$ have to have
following periodicity conditions
\begin{equation}
\tphi^i(2\pi)-\tphi^i(0)=2\pi n_i \  ,  
n_i \in Z \ . 
\end{equation}
Then integrating 
(\ref{relphitphi}) we obtain
the relation between the original
variables 
\begin{eqnarray}
\phi^1(2\pi)-\phi^1(0)&=&
2\pi(n_1+\gamma J_2) \ , \nonumber \\
\phi^2(2\pi)-\phi^2(0)&=&2\pi(n_2-\gamma J_1) \ , 
\nonumber \\
\end{eqnarray}
where 
\begin{equation}
J_i=\frac{1}{2\pi}\int_0^{2\pi}
d\sigma J^\tau_i(\sigma) \ ,
\end{equation}
and where $J_i$ are constant as follows
from (\ref{paju}). 

Now we can also look on this
problem from another point of view
using the fact that the momentum conjugate
to $\phi^i$ coincides with $J_i^\tau$. Therefore
we can rewrite the time
component of (\ref{relphitphi}) in the
form
\begin{equation}\label{relphitphi1}
\tilde{p}_i=p_i \ , \quad 
\partial_\sigma \tphi^i-
\partial_\sigma \phi^i=-\gamma_{ij}
p^j \ , \quad
i,j=1,\dots, d \ , 
\end{equation}
where $\gamma_{ij}=-\gamma_{ji}$ with
one nonzero component $\gamma_{12}\equiv\gamma$.
It is clear that (\ref{relphitphi1})
up to twisted boundary conditions a TsT 
transformation is just a simple linear
canonical transformation of the $U(1)$
isometry variables. Then the twist is the
origin of the nonequivalence of the original
and transformed theories. It is also clear
that the most general multi-parameter 
TsT transformed background obtained 
by applying TsT transformations
successively, many times when each time
we pick up a new torus and a new deformation
parameter is completely parametrised
by the relation (\ref{relphitphi1}) with
arbitrary matrix $\gamma_{ij}$. Therefore 
background that contains $d$ dimensional
torus admits $d(d-1)/2$ -parameter 
TsT transformation. In case of $AdS_5$
the most general TsT transformation 
applies to the five sphere $S^5$
(In order to preserve an isometry of
$AdS_5$)  has three independent parameters.
The twisted boundary conditions for the original
angles $\phi^i$ in the case of
the most general deformation take the form
\begin{equation}
\phi^i(2\pi)-\phi^i(0)=2\pi(n^i-\nu^i) \  ,
\quad 
\nu_i=-\gamma_{ik}J_k \ .
\end{equation}
The general three-parameter
$\gamma$-deformed background
is obtained by applying the
TsT transformation three
times. Following 
\cite{Alday:2005ww}
we express the corresponding
procedure as
\begin{equation}
(\phi_1,\phi_2,\phi_3)\stackrel{\gamma_3}{\rightarrow}
(\tilde{\phi}_1,\tilde{\phi}_2,\tilde{\phi}_3)
\stackrel{\gamma_1}{\rightarrow}
(\tilde{\tilde{\phi}}_1,\tilde{\tilde{\phi}}_2,\tilde{\tilde{\phi}}_3)
\stackrel{\gamma_2}{\rightarrow}
(\check{\phi}_1,\check{\phi}_2,\check{\phi}_3) \, .
\end{equation}
Since under every step
the corresponding Noether
currents remain the same
we can write
\begin{equation}
\begin{array}{lll}
\tilde{\phi}'_1-\phi'_1=-\gamma_3
J^{\tau}_2\,  ~~~&~~~ \tilde{\tilde{\phi}}'_1-\tilde{\phi}'_1=0\,
~~~&~~~ \check{\phi}_1-\tilde{\tilde{\phi}}'_1=\gamma_2 J^{\tau}_3\,
\\
\tilde{\phi}'_2-\phi'_2=\gamma_3 J^{\tau}_1 \,  ~~~&~~~
\tilde{\tilde{\phi}}'_2-\tilde{\phi}'_2=-\gamma_1 J^{\tau}_3 \,
~~~&~~~ \check{\phi}'_2 -\tilde{\tilde{\phi}}'_2=0\,
\\
\tilde{\phi}'_3-\phi'_3=0\, ~~~&~~~
\tilde{\tilde{\phi}}'_3-\tilde{\phi}'_3=\gamma_1 J^{\tau}_2 \,
~~~&~~~ \check{\phi}'_3-\tilde{\tilde{\phi}}'_3=-\gamma_2
J^{\tau}_1 \,
\end{array}
\end{equation}
From these formula's we can find
the relation between 
$\phi^i$ and $\tphi^i$ in the form
\begin{equation}
\partial_\sigma \tphi^i-
\partial_\sigma \phi^i=
\epsilon_{ijk}\gamma_j J^\tau_k \ ,
\quad 
\gamma_{ik}=-\epsilon_{ijk}\gamma_j
 \ . 
\end{equation}
Integrating this equation and 
using the fact that 
$\tphi^i(2\pi)-\tphi^i(0)=2\pi n^i$ we obtain the
twisted boundary conditions
for the original angles
\begin{equation}
\phi^i(2\pi)-\phi^i(0)=2\pi(n^i-\nu^i) \ , \quad 
\nu^i=\epsilon_{ijk}
\gamma_j J_k \ . 
\end{equation}
\section{Pure spinor action in 
$AdS_5\times S^5$ and explicit
coset representation}\label{third}
The pure spinor action in $AdS_5\times
S^5$ was introduced in  
\cite{Berkovits:2000yr,Berkovits:2004xu}
and further studied in 
\cite{Kluson:2006wq,Vallilo:2003nx}.
 In the covariant
worldsheet description the
pure spinor string action 
 on $AdS_5\times S^5$ takes the form
\begin{eqnarray}\label{Minaction}
S&=&-\frac{\sqrt{\lambda}}{2\pi}\int
d\tau d\sigma\sqrt{-\eta}\str[\frac{1}{2}
\eta^{\mu\nu}\left(J_\mu^{(2)}
J_{\nu}^{(2)}+J_\mu^{(1)}J_\nu^{(3)}
+J_\mu^{(3)}J_\nu^{(1)}\right)+
\nonumber\\
&+&\frac{\epsilon^{\mu\nu}}{4}
\left(J^{(1)}_\mu J^{(3)}_\nu-
J^{(3)}_\mu J^{(1)}_\nu\right)]+S_{ghost} \ , 
\nonumber \\
S_{ghost}&=&-\frac{\sqrt{\lambda}}{2\pi}
\int d\tau d\sigma \sqrt{-\eta}
\str[w_{\mu}\cP^{\mu\nu}
\partial_\nu \lambda
+\hw_{\mu}\tP^{\mu\nu}
\partial_\nu\hlambda
+\nonumber \\
&+&N_{\mu}
\cP^{\mu\nu}J^{(0)}_\nu+
\hN_{\mu}
\tP^{\mu\nu}J^{(0)}_\nu-
\frac{1}{2}N_{\mu}\cP^{\mu\nu}\hN_\nu
-\frac{1}{2}\hN_{\mu}\tP^{\mu\nu}
N_\nu] \ , 
\nonumber \\
\end{eqnarray}
where we have introduced the notation
\begin{eqnarray}
J^{(0)}_\mu&=&(g^{-1}\partial_\mu g)^{[\uc\ud]}T_{[\uc\ud]} \ ,
\quad J^{(1)}_\mu=(g^{-1}\partial_\mu g)^{\alpha}T_{\alpha} \ ,
\nonumber \\
J^{(2)}_\mu&=&(g^{-1}\partial_\mu g)^{\uc}T_{\uc} \ ,
\quad J^{(3)}_\mu=(g^{-1}\partial_\mu g)^{\halpha}
T_{\halpha} \ , \nonumber \\
w_\mu&=&w_{\mu\alpha} T_{\halpha}
\delta^{\alpha\halpha} \ , \quad 
\lambda=\lambda^\alpha T_\alpha \ ,
\nonumber \\
N_\mu&=&-\pb{w_\mu,\lambda}
=
-N_{\mu}^{cd}T_{[cd]}+N_\mu^{c'd'}T_{[c'd']}
 \ ,
\nonumber \\
\hw_\mu&=&\hw_{\mu\halpha} T_{\alpha}
\delta^{\halpha\alpha} \ , \quad 
\hlambda=\hlambda^{\halpha} T_{\halpha}  \ ,
 \nonumber \\
\hat{N}_\mu&=&-\pb{\hw_\mu,\hlambda}
=-\hN_{\mu}^{cd}T_{[cd]}+\hN_\mu^{c'd'}T_{[c'd']}
 \ . \nonumber \\
\end{eqnarray}
We also work with the flat
worldsheet metric where
$h_{\mu\nu}=\eta_{\mu\nu}=\mathrm{diag}(-1,1)$
and where we have also defined 
\begin{equation}
\cP^{\mu\nu}=\left(\eta^{\mu\nu}-\epsilon^{\mu\nu}
\right) \ , \quad
\tP^{\mu\nu}=\left(\eta^{\mu\nu}+\epsilon^{\mu\nu}
\right) \ , \quad  \epsilon^{\mu\nu}=
\frac{e^{\mu\nu}}{\sqrt{-\eta}} \ , \quad 
e^{01}=-e^{10}=1 \ .
\end{equation}
In what follows we work in 
coordinates $x^0=\tau, x^1=\sigma$
where $\sigma\in (0,2\pi)$. 

An element $M$ of the superalgebra
$\mathbf{su}(2,2|4)$ is given by a $8\times 8$
matrix that can be written
in terms of $4\times 4$ blocks
as
\begin{equation}\label{Gsup}
M=\left(\begin{array}{cc}
A & X \\
Y & D \\
\end{array}\right) \ . 
\end{equation}
The superalgebra $\mathbf{su}(2,2|4)$ is
singled out by requiring that
$M$ has to have zero supertrace
$\str M=\tr A-\tr D=0$ and to satisfy
the following reality condition
\begin{equation}\label{HM}
HM+M^{\dag}H=0 \ .
\end{equation}
The choice of the hermitian matrix $H$ is
not unique and we choose $H$ to be
of the diagonal form
\begin{equation}\label{Hdef}
H=\left(\begin{array}{cc}
\Sigma & 0 \\
0 & 1 \\ \end{array}\right) \ . 
\end{equation}
Then (\ref{HM}) and (\ref{Hdef})
imply
 \begin{equation}
D=-D^{\dag} \ , \quad 
 \Sigma A=-A^{\dag}\Sigma \ , \quad
Y=-X^{\dag}\Sigma \ ,
\end{equation}
where
\begin{equation}
\Sigma=\left(\begin{array}{cccc}
1 & 0 & 0 & 0 \\
0 & 1 & 0 & 0 \\
0 & 0 & -1 & 0 \\
0 & 0 & 0 & -1 \\
\end{array}\right) \ . 
\end{equation}
The algebra $\mathbf{su}(2,2|4)$ also
contains the $\mathbf{u}(1)$ generator
$i\mathbf{I}$ where $\mathbf{I}$ is identity
matrix of the corresponding dimension. 
The superalgebra $\mathbf{psu}(2,2|4)$ is defined
as the quotient algebra of $\mathbf{su}(2,2|4)$
over this $\mathbf{u}(1)$ factor; it has
no realisation in terms of $8\times 8$ matrices.
 
The essential feature of the superalgebra
$\mathbf{su}(2,2|4)$ is that it admits
a $\mathbf{Z}_4$ automorphism $\Omega$ 
such that the condition $\Omega(H)=H$
determines the maximal subgroup
to be $SO(4,1)\times SO(5)$ that
leads to the definition of the
coset for the sigma model.
The $\mathbf{Z}_4$ automorphism
$\Omega$ takes an element of $\mathbf{psu}(2,2|4)$
to another $G\rightarrow \Omega(G)$
such that
\begin{equation}\label{Omegadef}
\Omega(G)=\left(\begin{array}{cc}
KA^TK & -KY^TK  \\
KX^TK & KB^TK \\ \end{array}
\right)  \ , \quad 
K=\left(\begin{array}{cccc}
0 & 1 & 0 & 0 \\
-1 & 0 & 0 & 0\\
0 & 0 & 0 & 1 \\
0 & 0 & -1 & 0 \\
\end{array}\right) \ .
\end{equation}
Since $\Omega^4=1$ the eigenvalues of
$\Omega$ are $i^p \ , p=0,1,2,3$. Therefore
we can decompose the superalgebra
$G$ as
\begin{equation}
G=\mH_0\oplus\mH_1\oplus\mH_2\oplus \mH_3 \ ,
\end{equation}
where $\mH_p$ denotes the eigenspace of
$\Omega$  such that if $H\in \mH_p$ then
\begin{equation}\label{Omegaav}
\Omega(H)=i^pH \ .
\end{equation}
Explicitly, any matrix $M$ of $\mathbf{su}(2,2|4)$
can be decomposed into the sum
\begin{equation}
M=M^{(0)}+M^{(1)}+M^{(2)}+M^{(3)} \ ,
\end{equation}
where 
\begin{eqnarray}
M^{(0)}&=&\frac{1}{4}
(M+\Omega(M)+\Omega^2(M)+\Omega^3(M))=\frac{1}{2}
\left(\begin{array}{cc}
A+KA^TK & 0 \\
0 & D+KD^T K \\ \end{array}
\right) \ , \nonumber \\
M^{(2)}&=&\frac{1}{4}
(M-\Omega(M)+\Omega^2(M)-\Omega^3(M))=\frac{1}{2}
\left(\begin{array}{cc}
A-KA^TK & 0 \\
0 & D-KD^T K \\ \end{array}
\right) \ , \nonumber \\
M^{(1)}&=&\frac{1}{4}
(M-i\Omega(M)-\Omega^2(M)+i\Omega^3(M))=\frac{1}{2}
\left(\begin{array}{cc}
 0 & X+iKY^TK \\
Y-iKX^TK & 0  \\ \end{array}
\right) \ , \nonumber \\
M^{(3)}&=&\frac{1}{4}
(M+i\Omega(M)-\Omega^2(M)-i\Omega^3(M))=\frac{1}{2}
\left(\begin{array}{cc}
 0 & X-iKY^TK \\
Y+iKX^TK & 0  \\ \end{array}
\right) \ , \nonumber \\
\end{eqnarray}
and where  $\Omega(M^{(p)})=i^p M^{(p)} $.
We see that $M^{(0)}$ form
$\mathbf{so}(4,1)\times 
\mathbf{so}(5)$ subalgebra which
we wish to mod out in the coset.
We also see that the matrices $M^{(1,3)}$
contain the odd matrices. Splitting $M$
into Grassman  even and odd parts
\begin{equation}
M=M_{even}+M_{odd}
\ , \quad 
M_{even}=\left(\begin{array}{cc}
A & 0 \\
0 & D \\ \end{array}\right) \ , \quad 
M_{odd}=\left(\begin{array}{cc}
0 & X \\
Y & 0 \\ \end{array}\right) \ ,
\end{equation}
we can rewrite the expressions
for $M^{(p)}$ in the following
form
\begin{eqnarray}\label{Mievenodd}
M^{(0)}&=&\frac{1}{2}(M_{even}+K_8M_{even}^T
K_8) \ , \quad 
M^{(2)}=\frac{1}{2}(M_{even}-K_8M^T_{even}K_8) \ ,
\nonumber \\
M^{(1)}&=&\frac{1}{2}(M_{odd}+
i\tilde{K}_8 M^T_{odd}K_8) \ ,
\quad 
M^{(3)}=\frac{1}{2}
(M_{odd}-i\tilde{K}_8 M^T_{odd}
K_8) \ ,
\nonumber \\
\end{eqnarray}
where $K_8$ and $\tilde{K}_8$ are
defined as
\begin{equation}
K_8=\left(\begin{array}{cc}
K & 0 \\
0 & K \\ \end{array}\right) \ ,
\quad
\tilde{K}_8=\left(\begin{array}{cc}
K & 0 \\
0 & -K \\ \end{array}\right) \ .
\end{equation}
The next step is to explicit
choose the coset representative $g$.
Following \cite{Alday:2005ww}
we take
the coset parametrisation in the form
\begin{equation}\label{cosrep}
g=g(\theta)g(z) \ .
\end{equation}
Here $g(\theta)=\exp(\theta)$, where
$\theta$ is an  element of $\mathbf{psu}(2,2|4)$
that contains $32$ fermionic degrees
of freedom. 
The element $g(z)$ 
belongs to $SU(2,2)
\times SU(4)$ 
and takes following form
\cite{Alday:2005gi}
\begin{equation}\label{gze}
g(z)=\left(\begin{array}{cc}
\tg_a(x) & 0 \\
0 & \tg_s(y) \\ \end{array}\right) \ , 
\end{equation}
where
\begin{equation}
\tg_a(x)=\exp\frac{1}{2}(x_a\gamma_a) \ , \quad
\tg_s(y)=\exp\frac{\mathrm{i}}{2}(y_a\Gamma_a) 
 \ ,
\end{equation} 
where $z\equiv(x_a,y_a)$ and $x_a$
parametrise the $AdS_5$ space while
$y_a$ stand for the five-sphere. 
The matrices $\Gamma_a,\gamma_a \ , a=1,\dots,5$ 
are Dirac matrices for $SO(5)$ and
$SO(4,1)$ respectively. These matrices
obey the relations
\begin{equation}\label{KgammaK}
K\Gamma_a^TK=-\Gamma_a \ , \quad  
K\gamma_a^TK=-\gamma_a \ .
\end{equation}
 Using this property of the Dirac matrices
it can be easily shown that they span
the orthogonal complements to the Lie
algebras $\mathbf{so}(5)$ and $\mathbf{so}(4,1)$ 
respectively
\footnote{For very nice discussion, see
\cite{Frolov:2006cc}.}.
Now with the choice of the
 coset representative
given in (\ref{cosrep})  the current
takes the form
\begin{equation}\label{Jthetaz}
J=g^{-1}dg=g^{-1}(z)g^{-1}(\theta)
dg(\theta)g(z)+g^{-1}(z)dg(z) \ .
\end{equation}
Since 
\begin{eqnarray}
g(\theta)=
\cosh\theta+\sinh\theta \ , \quad 
g^{-1}(\theta)&=&\cosh\theta-\sinh\theta \  
\nonumber \\
\end{eqnarray}
we get
\begin{equation}\label{BF}
g^{-1}(\theta)dg(\theta)=(\cosh\theta-\sinh\theta)
(d\cosh\theta+d\sinh\theta)=F+B \ , 
\end{equation}
where
\begin{eqnarray}
B&\equiv&\cosh\theta d\cosh\theta-\sinh
\theta d\sinh\theta \ , \nonumber \\
F&\equiv&\cosh\theta d\sinh\theta
-\sinh\theta d\cosh\theta \ 
\nonumber \\  
\end{eqnarray}
are  even (contain even
number of $\theta$'s) and
odd (contain odd number of
$\theta$'s) element respectively.
 With the help
(\ref{Jthetaz}) and  (\ref{BF}) we
find that the  even   component of $J$ 
takes the  form 
\begin{equation}\label{Jeven}
J_{even}=g^{-1}(z)Bg(z)+g^{-1}(z)dg(z) 
\end{equation}
while the odd element is equal
to \begin{equation}\label{Jodd}
J_{odd}=g^{-1}(z)Fg(z) \ .
\end{equation}
As the next step we 
 find
components of the current
$J^{(i)}$ that belongs
to appropriate subspaces
$\mH^{(i)}$.
To do this we use the
relation (\ref{Mievenodd}).
To present further result we
 define
\begin{equation}\label{defG}
G=g(z)K_8g^t(z)=\left(\begin{array}{cc}
g_a  & 0 \\
0 & g_s \\ \end{array}
\right) \ , \quad 
\tG=g(z)\tilde{K}_8g^t(z)=\left(\begin{array}{cc}
g_a  & 0 \\
0 & -g_s \\ \end{array}
\right) \ . 
\end{equation}
As was argued in \cite{Alday:2005ww}
the $4\times 4$ matrices $g_a
\in SU(2,2)$ and $g_s\in
SU(4)$ provide another
parametrisation of
the five-sphere and the AdS space.
On coordinates $z$ the global symmetry
algebra is  realised
non-linearly. In opposite, 
$g_a$ and $g_s$ carry a linear
representation of the superconformal
algebra. 

Now using (\ref{Mievenodd}) and
(\ref{defG}) we obtain
\begin{eqnarray}\label{J0gb}
2J^{(0)}=J_{even}+K_8 J_{even}^T
K_8
=2g^{-1}dg+
g^{-1}(B-GB^TG^{-1}-dGG^{-1})g \ 
\nonumber \\
\end{eqnarray}
 using the fact that
$K^{-1}_8=-K_8 \ , 
\tilde{K}^{-1}_8=-\tilde{K}^{-1}_8$.
In (\ref{J0gb}) $g$ means 
$g(z)$ and in the following we
use this notation.  
In the same way we obtain
\begin{equation}\label{J2gb}
2J^{(2)}=J_{even}-K_8 J_{even}^T
K_8=g^{-1}(B+GB^TG^{-1}+dGG^{-1})g \ 
 \end{equation}
and
\begin{eqnarray}\label{J1gb}
2J^{(1)}&=&
J_{odd}+i\tilde{K}_8 J^T_{odd}K_8
=g^{-1}(F-i\tG F^TG^{-1})g \ ,
\nonumber \\
2J^{(3)}&=&J_{odd}-i\tilde{K}_8
J^T_{odd}K_8=g^{-1}(F+i\tG F^T G^{-1})g \ . 
\nonumber \\
\end{eqnarray}
With the help of (\ref{J0gb}),
(\ref{J2gb}) and (\ref{J1gb})
 we can
write the pure spinor Lagrangian
density  in the form
\begin{eqnarray}\label{Lcoset}
\mL&=&-\frac{\sqrt{\lambda}}{8\pi}
\str[\frac{1}{2}
\eta^{\mu\nu}
(B_\mu+GB^T_\mu G^{-1}+\partial_\mu
GG^{-1})
(B_\nu+GB^T_\nu G^{-1}
+\partial_\nu GG^{-1})
+\nonumber \\
&+&\eta^{\mu\nu}(F_\mu-i\tG F^T_\mu G^{-1})
(F_\nu+i\tG F^T_\nu G^{-1})
+\frac{\epsilon^{\mu\nu}}{2}
(F_\mu-i\tG F^T_\mu G^{-1})
(F_\nu+i\tG F^T_\nu G^{-1})]
\nonumber \\
&-&\frac{\sqrt{\lambda}}{2\pi}
\str(w_{\mu}\cP^{\mu\nu}
\partial_\nu \lambda
+\hw_{\mu}\tP^{\mu\nu}
\partial_\nu\hlambda
-N_{\mu}\cP^{\mu\nu}\hN_\nu
+\nonumber \\
&+&\frac{1}{2}N_{\mu}
\cP^{\mu\nu}
(2g^{-1}\partial_\nu g+
g^{-1}(B_\nu-
GB^T_\nu G^{-1}-\partial_\nu
GG^{-1})g)+
\nonumber \\
&+&\frac{1}{2}\hN_{\mu}
\tP^{\mu\nu}
(2g^{-1}\partial_\nu g+
g^{-1}(B_\nu-
GB^T_\nu G^{-1}-\partial_\nu
GG^{-1})g)) \ . 
\nonumber \\
\end{eqnarray}
By using the cyclic property
of the supertrace the action
can be further simplified
using the fact that
\begin{eqnarray}
\str\tG F^T_\mu G^{-1}\tG F_\nu^T
G^{-1}=\str F^T_\mu \left(\begin{array}{cc}
1 & 0 \\
0 & -1 \\ \end{array}\right)
F_\nu^T 
\left(\begin{array}{cc}
1 & 0 \\
0 & -1 \\ \end{array}\right)
=\str F_\mu F_\nu\ ,
\nonumber\\
\end{eqnarray}
where we have used 
\begin{equation}
G^{-1}\tG=\left(\begin{array}{cc}
1 & 0 \\
0 & -1 \\ \end{array}\right)
\end{equation}
and also the fact that
 that $F$ is off-diagonal matrix.
Then we can simplify the action
(\ref{Lcoset}) as 
\begin{eqnarray}\label{LcosetS}
\mL&=&-\frac{\sqrt{\lambda}}{8\pi}
\str[\frac{1}{2}
\eta^{\mu\nu}
(B_\mu+GB^T_\mu G^{-1}+\partial_\mu
GG^{-1})
(B_\nu+GB^T_\nu G^{-1}
+\partial_\nu GG^{-1})
+\nonumber \\
&+&2\eta^{\mu\nu}(F_{\mu}F_\nu-i\tG F^T_\mu G^{-1}F_\nu)
+\epsilon^{\mu\nu}(
F_{\mu}F_{\nu}-i\tG F^T_\mu G^{-1}
F_\nu)]-
\nonumber \\
&-&\frac{\sqrt{\lambda}}{2\pi}
\str[w_{\mu}\cP^{\mu\nu}
\partial_\nu \lambda
+\hw_{\mu}\tP^{\mu\nu}
\partial_\nu\hlambda
-N_{\mu}\cP^{\mu\nu}\hN_\nu
+\nonumber \\
&+&\frac{1}{2}N_{\mu}
\cP^{\mu\nu}
(2g^{-1}\partial_\nu g+
g^{-1}(B_\nu-
GB^T_\nu G^{-1}-\partial_\nu
GG^{-1})g)+
\nonumber \\
&+&\frac{1}{2}\hN_{\mu}
\tP^{\mu\nu}
(2g^{-1}\partial_\nu g+
g^{-1}(B_\nu-
GB^T_\nu G^{-1}-\partial_\nu
GG^{-1})g)] \ . 
\nonumber \\
\end{eqnarray}
We see that
the pure spinor parts of
the action is rather  complicated.
In fact, the presence of
the matrix $g$ makes the analysis
 difficult since the
symmetries do not act on it
linearly. To resolve this
problem  we begin with the observation
that 
\begin{eqnarray}
&\str& (N_\mu \cP^{\mu\nu}J_\nu^{(0)})=
-\str(\pb{w_\mu,\lambda}\cP^{\mu\nu}
J^{(0)}_\nu)= \nonumber \\
&=&-\cP^{\mu\nu}
\str(w_\mu (\lambda J^{(0)}_\nu-J^{(0)}_\nu \lambda))
=\str w_\mu \cP^{\mu\nu}\com{J^{(0)}_\nu,\lambda} \ , 
\nonumber \\
\end{eqnarray}
where we have used 
 the fact that the off-diagonal
blocks of the matrices $w,\lambda$  contain
Grassman even elements. In the same we can proceed
with $\hN$ and then we can 
rewrite  the pure
spinor Lagrangian into the form
\begin{eqnarray}\label{pureg}
\mL_{pure}=-\frac{\sqrt{\lambda}}{2\pi}
\str (w_\mu \cP^{\mu\nu}\nabla_\nu \lambda+
\hw_\mu \tP^{\mu\nu}\nabla_\nu \hlambda
-N_\mu \cP^{\mu\nu}\hN_\nu) \ , 
\nonumber \\
\end{eqnarray} 
where 
\begin{equation}
\nabla_\mu X=\partial_\mu X+[J^{(0)}_\mu,
X] \ . 
\end{equation}
The form of the current $J^{(0)}$
given in (\ref{J0gb}) suggests the
following  field redefinition of
the ghost variables 
\begin{eqnarray}\label{lambdare}
\lambda &=&g^{-1}\olambda g \ , \quad
\hlambda=g^{-1}\uhlambda g  \ , \nonumber \\
w_\mu&=&g^{-1}\ow_\mu g \ ,  
\quad 
\hw_\mu= g^{-1}\uhw_\mu g \ . 
\nonumber \\
\end{eqnarray}
First of all we have to check that 
the new pure spinors matrices $\olambda \ ,
\uhlambda$ still belong to $M^{(1)},M^{(3)}$
respectively. To do this we use the
fact that $\lambda$ has schematically following
form
\begin{equation}
\lambda=\left(\begin{array}{cc}
0 & X_\lambda \\
Y_\lambda  & 0 \\ \end{array}   
\right)
\end{equation} 
and hence 
\begin{equation}\label{relambda}
\olambda=g\lambda g^{-1}=\left(\begin{array}{cc}
0 & \tg_a^{-1}X_\lambda \tg_s\\
\tg_s^{-1}Y_\lambda \tg_a  & 0 \\ 
\end{array} \right) \ , 
\end{equation}
where we have used the explicit form of the coset
element given in (\ref{gze}). Then using the
 definition of $\Omega$ given in 
(\ref{Omegadef}) we get
\begin{eqnarray}
\Omega(\olambda)&=&
\left(\begin{array}{cc}
0 & -K(\tg_s^{-1}Y_\lambda \tg_a)^T K  \\
K(\tg_a^{-1}X_\lambda \tg_s)^TK & 0 \\ 
\end{array}\right)\nonumber \\
&=&\left(\begin{array}{cc}
\tg_a & 0 \\
0 & \tg_s \\ \end{array}\right)
\left(\begin{array}{cc}
0 & -KY_\lambda^T K  \\
KX_\lambda^TK & 0 \\ 
\end{array}\right)
\left(\begin{array}{cc}
\tg_a^{-1} & 0 \\
0 & \tg_s^{-1} \\ \end{array}\right)
=g\Omega(\lambda)g^{-1}  \ . 
\nonumber \\
\end{eqnarray}
Since $\Omega(\lambda)=i\lambda$ the
equation above implies
\begin{equation}
\Omega(\olambda)=ig \lambda g^{-1}=i\olambda
\end{equation}
and hence $\olambda$ belongs to $M^{(1)}$
as well. In the same way we can
show that $\uhlambda$ belongs to
$M^{(3)}$ and hence the field redefinition 
(\ref{relambda}) is well defined. 

It is easy to see that if the original pure
spinors $\lambda,\hlambda$ 
obey the pure spinor conditions then 
$\olambda,\uhlambda$ obey these conditions
as well. More precisely, 
note that the pure spinor condition for 
$\lambda$ can be written as
\begin{equation}\label{purespinm}
\pb{\lambda,\lambda}=
\lambda^\alpha\lambda^\beta 
\pb{T_\alpha,T_\beta}=
\lambda^\alpha\lambda^\beta f_{\alpha\beta}^{\uc}
T_{\uc}\sim  \lambda^\alpha \gamma_{\alpha\beta}^{\uc}
\lambda^\beta
T_{\uc}=0 \ . 
\end{equation}
Then if we insert (\ref{lambdare}) into
(\ref{purespinm}) we easily get
\begin{equation}
\pb{g^{-1}\olambda g,g^{-1}\olambda
g}=g^{-1}\pb{\olambda,\olambda}g=0
\Rightarrow
\pb{\olambda,\olambda}=0 \ 
\end{equation}
so that $\olambda$ obey the pure
spinor constraint as well. It is clear
that the same analysis can be performed
for $\hlambda$ as well and we obtain
that $\uhlambda$ obey the pure spinor
conditions.
Now with the help of 
(\ref{lambdare}) we obtain 
\begin{eqnarray}
w_\mu \cP^{\mu\nu}\nabla_\nu \lambda&=&
g^{-1}\ow_\mu \cP^{\mu\nu}
(\partial_\nu \olambda+
\frac{1}{2}\com{(B-GB^TG^{-1}-dGG^{-1}) ,\olambda})g
\nonumber \\
\hw_\mu \tP^{\mu\nu}\nabla_\nu \hlambda&=&
g^{-1}\uhw_\mu \tP^{\mu\nu}
(\partial_\nu \uhlambda+
\frac{1}{2}\com{(B-GB^TG^{-1}-dGG^{-1}) ,\uhlambda})g
\nonumber \\
N_\mu&=&g^{-1}\overline{N}_\mu g \ , \nonumber \\
\hN_\mu&=&g^{-1}\underline{\hN}_\mu g \ . \nonumber \\
\end{eqnarray}
To simplify further analysis
we introduce following notation
\begin{eqnarray}\label{J0gb2}
J^{(0)}&=&
g^{-1}dg+\frac{1}{2}[
g^{-1}(B-GB^TG^{-1}-dGG^{-1})g]=
g^{-1}dg+g^{-1}\bJ^{(0)}g  \ ,
\nonumber \\
J^{(2)}&=&
\frac{1}{2}[g^{-1}(B+GB^TG^{-1}+dGG^{-1})g]
=g^{-1}\bJ^{(2)}g \ ,  
\nonumber \\
J^{(1)}&=&
\frac{1}{2}g^{-1}(F-i\tG F^TG^{-1})g
=g^{-1}\bJ^{(1)}g
 \ ,
\nonumber \\
J^{(3)}&=&\frac{1}{2}
g^{-1}(F+i\tG F^T G^{-1})g
=g^{-1}\bJ^{(3)}g \ . 
\nonumber \\
\end{eqnarray}
The we can  write the  pure
spinor action in the same
form as in (\ref{Minaction})
\begin{eqnarray}\label{LprF}
S&=&-\frac{\sqrt{\lambda}}{2\pi}\int
d\tau d\sigma\sqrt{-\eta}\str[\frac{1}{2}
\eta^{\mu\nu}\left(\bJ_\mu^{(2)}
\bJ_{\nu}^{(2)}+\bJ_\mu^{(1)}\bJ_\nu^{(3)}
+\bJ_\mu^{(3)}\bJ_\nu^{(1)}\right)+
\nonumber\\
&+&\frac{\epsilon^{\mu\nu}}{4}
\left(\bJ^{(1)}_\mu \bJ^{(3)}_\nu-
\bJ^{(3)}_\mu \bJ^{(1)}_\nu\right)]+S_{ghost} \ , 
\nonumber \\
S_{ghost}&=&-\frac{\sqrt{\lambda}}{2\pi}
\int d\tau d\sigma \sqrt{-\eta}
\str[\ow_{\mu}\cP^{\mu\nu}
\partial_\nu \olambda
+\uhw_{\mu}\tP^{\mu\nu}
\partial_\nu\uhlambda
+\nonumber \\
&+&\overline{N}_{\mu}
\cP^{\mu\nu}\bJ^{(0)}_\nu+
\underline{\hN}_{\mu}
\tP^{\mu\nu}\bJ^{(0)}_\nu-
\frac{1}{2}\overline{N}_{\mu}
\cP^{\mu\nu}\underline{\hN}_\nu
-\frac{1}{2}\underline{\hN}_{\mu}\tP^{\mu\nu}
\overline{N}_\nu] \ . 
\nonumber \\
\end{eqnarray}
However there is one crucial difference
between the action (\ref{LprF}) and
(\ref{Minaction}). Due to the explicit
form of the coset representative 
(\ref{cosrep}) it is clear that the
currents $\bJ^{(i)}$ do not transform under
the gauge transformations as the
original one $J^{(i)}$. More precisely, the original
action (\ref{Minaction}) was invariant
under the gauge transformations
\begin{eqnarray}\label{gauge}
J'&=&h^{-1}Jh+h^{-1}dh \ , 
\quad \lambda'=h^{-1}\lambda h \ ,  \quad
\hlambda'=h^{-1}\hlambda h \ , 
\nonumber \\
w'_\mu&=&h^{-1} w_\mu h \ , \quad 
\hw'_\mu=h^{-1}\hw_\mu h \ ,
\nonumber \\
\end{eqnarray}
where $h$ belongs to $SO(4,1)\times SO(5)$. 
Clearly the redefined currents 
$\bJ$ and ghost variables do not
transform in the same way as 
(\ref{gauge}).  This follows from 
the fact that  the choice
of given coset representative 
effectively fixes  given gauge symmetry.
For that reason the action (\ref{LprF})
does not possess the gauge symmetry of
the original action. 

We conclude this section with 
the brief discussion of the properties
of the matrix $G$. 
With the certain choice
of the matrix $K$ the matrix
$g_s$ parameterising $S^5$
can be written as follows
\begin{equation}
g_s=\left(\begin{array}{cccc}
0 & u_3 & u_1 & u_2 \\
-u_3 & 0 & u_2^* & -u_1^* \\
-u_1 & -u_2^* & 0 & u_3^* \\
-u_2 & u_1^* & -u_3^* & 0 \\
\end{array}\right) \ . 
\end{equation} 
This is unitary matrix
$g_s^\dag g_s=1$ on the condition
that the three complex coordinates
$u_i$ obey the constraint
$|u_1|^2+|u_2|^2+
|u_3|^2=1$. A similar parameterisation
of the $AdS_5$ space is given by
\begin{equation}
g_a=\left(\begin{array}{cccc}
0 & v_3 & v_1 & v_2 \\
-v_3 & 0 & -v_2^* & v_1^* \\
-v_1 & v_2^* & 0 & v_3^* \\
-v_2 & -v_1^* & -v_3^* & 0 \\
\end{array}\right) \ , 
\end{equation}
where now $g_a \in SU(2,2)$
so that it obeys
$g^\dag_a E g_a=E$ where
$E=\mathrm{diag}(1,1,-1,-1)$
provided the complex
numbers $v_i$ satisfy the constraint
$|v_1|^2+|v_2|^2-|v_3|^2=-1 $.
\section{Equation of motions and
BRST invariance}\label{fourth}
Our goal  is to express the equations
of motion that follow from the action
(\ref{Minaction}) in terms of 
redefined ghost fields
(\ref{lambdare}) and of  the
currents $\bJ^{(i)}$ defined in 
(\ref{J0gb2}). 
We firstly write the equations 
of motion that arise from (\ref{Minaction}).
These equations of motion were
determined previously in  
\cite{Berkovits:2000yr}
and their covariant formulation was 
also given in \cite{Kluson:2006wq}
\begin{eqnarray}\label{eqm}
\tP^{\mu\nu}
\nabla_\mu J^{(3)}_\nu
+
[J_\nu^{(3)},N_\mu]\cP^{\mu\nu}+
[J_\nu^{(3)},\hN_\mu]\tP^{\mu\nu}
=0 \ , \nonumber \\
\cP^{\mu\nu}\nabla_\mu J^{(1)}_\nu
+
[J_\nu^{(1)},N_\mu]\cP^{\mu\nu}+
[J_\nu^{(1)},\hN_\mu]\tP^{\mu\nu}
=0 \ , \nonumber \\
\cP^{\mu\nu}
\nabla_\mu J^{(2)}_\nu
-\epsilon^{\mu\nu}
[J^{(1)}_\mu,J^{(1)}_\nu]
+[J_\nu^{(2)},N_\mu]\cP^{\mu\nu}+
[J_\nu^{(2)},\hN_\mu]\tP^{\mu\nu}
=0 \ , \nonumber \\
\tP^{\mu\nu}
\nabla_\mu J^{(2)}_\nu
+\epsilon^{\mu\nu}
[J^{(3)}_\mu,J^{(3)}_\nu]
+[J_\nu^{(2)},N_\mu]\cP^{\mu\nu}+
[J_\nu^{(2)},\hN_\mu]\tP^{\mu\nu}
=0 \ ,
\nonumber \\
\cP^{\mu\nu}\nabla_\nu\lambda+
\cP^{\mu\nu}[\lambda,\hN_\nu]
=0 \ ,
\nonumber \\
\tP^{\mu\nu}\nabla_\nu \hlambda
+\tP^{\mu\nu}[\hlambda,N_\nu]
=0 \ ,\nonumber \\
\end{eqnarray}
where
\begin{eqnarray}
\nabla_\nu J^{(i)}_\mu&=&
\partial_\nu J^{(i)}_\mu+
[J^{(0)}_\nu,J^{(i)}_\mu] \ ,
\nonumber \\
\nabla_\mu \lambda&=&
\partial_\mu\lambda+\com{J^{(0)}_\mu,
 \lambda} \ .
\nonumber \\
\end{eqnarray}
Now we rewrite these equations
of motion using the 
form of the currents
given in (\ref{J0gb2}) and
we obtain
\begin{eqnarray}
\nabla_\mu J^{(i)}_\nu &=&
g^{-1}(\partial_\mu \bJ_\nu^{(i)}+
\com{\bJ^{(0)}_\mu,J^{(i)}_\nu})g
\equiv g^{-1}\onabla_\nu J^{(i)}_\mu
g \ , \quad  i=1,2,3 \ , 
 \nonumber \\
\nabla_\mu \lambda &=&
g^{-1}\onabla_\mu \olambda g \ , \quad  
\nabla_\mu \hlambda= g^{-1} \onabla_\mu \uhlambda g  \ . 
\nonumber \\
\end{eqnarray}
Then it is easy to see that the
equations of motion given above 
take the form
\begin{equation}\label{eqm1r}
\tP^{\mu\nu}
\onabla_\mu \bJ^{(3)}_\nu
+
[\bJ_\nu^{(3)},\overline{N}_\mu]\cP^{\mu\nu}+
[\bJ_\nu^{(3)},\underline{\hN}_\mu]\tP^{\mu\nu}
=0 \ ,
\end{equation}
\begin{equation}\label{eqm2r}
\cP^{\mu\nu}\onabla_\mu \bJ^{(1)}_\nu
+
[\bJ_\nu^{(1)},\overline{N}_\mu]\cP^{\mu\nu}+
[\bJ_\nu^{(1)},\underline{\hN}_\mu]\tP^{\mu\nu}
=0 \ ,
\end{equation}
\begin{equation}\label{eqm3ar}
\cP^{\mu\nu}
\onabla_\mu \bJ^{(2)}_\nu
-\epsilon^{\mu\nu}
[\bJ^{(1)}_\mu,\bJ^{(1)}_\nu]
+[\bJ_\nu^{(2)},\overline{N}_\mu]\cP^{\mu\nu}+
[\bJ_\nu^{(2)},\underline{\hN}_\mu]\tP^{\mu\nu}
=0 \ ,
\end{equation}
\begin{equation}\label{eqm3br}
\tP^{\mu\nu}
\onabla_\mu \bJ^{(2)}_\nu
+\epsilon^{\mu\nu}
[\bJ^{(3)}_\mu,\bJ^{(3)}_\nu]
+[\bJ_\nu^{(2)},\overline{N}_\mu]\cP^{\mu\nu}+
[\bJ_\nu^{(2)},\underline{\hN}_\mu]\tP^{\mu\nu}
=0 \ ,
\end{equation}
\begin{equation}\label{eqg1r}
\cP^{\mu\nu}\onabla_\nu\olambda+
\cP^{\mu\nu}[\olambda,\underline{\hN}_\nu]
=0 \ ,
\end{equation}
\begin{equation}\label{eqg2r}
\tP^{\mu\nu}\onabla_\nu \uhlambda
+\tP^{\mu\nu}[\uhlambda,\overline{N}_\nu]
=0 \ .
\end{equation}
The fact that in the new
variables the equations of motion
have the same form as 
the equations given in (\ref{eqm})  
has an important consequence for
the conservation of the BRST currents.
These currents are defined as  
\cite{Berkovits:2000yr}
\begin{equation}\label{defr}
j^\mu_R=\str(\hlambda J_\nu^{(1)}
\tP^{\nu\mu}) \ , \quad 
j^\mu_L=\str(\lambda J_\nu^{(3)}
\cP^{\nu\mu})
\end{equation}
and  they are conserved
\begin{equation}\label{djrl}
\partial_\mu j^\mu_{R,L}=0 \ .
\end{equation}
With the help of
(\ref{lambdare}) and 
 (\ref{J0gb2}) we can 
rewrite (\ref{defr}) into the
form
\begin{equation}\label{jrl}
j^\mu_R=\str(\uhlambda \bJ_\nu^{(1)}
\tP^{\nu\mu}) \ , \quad 
j^\mu_L=\str(\olambda \bJ_\nu^{(3)}
\cP^{\nu\mu}) \ .  
\end{equation}
Now we will show that these currents
are conserved as well. To do this
we will calculate  
$\partial_\mu j^\mu_L$ using the equations of motion
 (\ref{eqm1r}) and (\ref{eqg1r})
\begin{eqnarray}
\partial_\mu j^\mu_L&=&
\str (\cP^{\nu\mu}
\partial_\mu \olambda \bJ^{(3)}_\nu)+
\str (\olambda \tP^{\mu\nu}\partial_\mu
\bJ^{(3)}_\nu)\nonumber \\
&=& \str((-\cP^{\nu\mu}[\uhlambda,\underline{\hN}_\mu]
-\cP^{\mu\nu}[\bJ^{(0)}_\mu,\olambda])
\bJ^{(3)}_\nu)+\nonumber \\
&+&\str(\olambda(-\tP^{\mu\nu}[\bJ^{(0)}_\mu,\bJ^{(3)}_\nu]
-[\bJ^{(3)}_\nu,\overline{N}_\mu]\cP^{\mu\nu}
-[\bJ_\nu^{(3)},\underline{\hN}_\mu]\tP^{\mu\nu})\nonumber \\
&=&
 -\str (\bJ^{(3)}_\nu
\com{\overline{N}_\mu,\olambda})
\cP^{\mu\nu}
\nonumber \\
\end{eqnarray}  
that vanishes thanks to the pure spinor
constraint. In the same way we can prove
the conservation of $j^\mu_R$. 
The existence of two conserved
BRST currents (\ref{jrl}) imply that we
can define two BRST charges  
\begin{equation}
Q_L=\frac{1}{2\pi}
\int_0^{2\pi}
d\sigma j_L^0 \ , \quad  
Q_R=\frac{1}{2\pi}
\int_0^{2\pi}
d\sigma j_R^0 \  .
\end{equation}
Let us now calculate their time
derivative  
\begin{eqnarray}\label{dtQ}
\frac{dQ_L}{d\tau}=-\frac{1}{2\pi}\int_0^{2\pi}
d\sigma \partial_\sigma j^1_L=-\frac{1}{2\pi}
(j^1_L(2\pi)-j_L^1(0)) \ , 
\nonumber \\
\frac{dQ_R}{d\tau}=-\frac{1}{2\pi}\int_0^{2\pi}
d\sigma \partial_\sigma j^1_R=-\frac{1}{2\pi}
(j^1_R(2\pi)-j_R^1(0)) \ , 
\nonumber \\
\end{eqnarray}
where we have used (\ref{djrl}). 
For ordinary closed string we
demand that the world-sheet
fields are periodic 
\begin{equation}
j^1_{L,R}(\tau,\sigma+2\pi)=j^1_{L,R}(\tau,\sigma) \ . 
\end{equation}
Then  (\ref{dtQ}) 
implies that $Q_L,Q_R$ are
time-independent. 
Even if these results are well known
we reviewed here since as we will see
in the next section the world-volume
modes do not have to be periodic and
hence the existence of the conserved
BRST charges is not generally obvious.
\section{Fermions and 
pure spinors twisting}\label{fifth}
The original fermions and pure spinors 
  transform  under isometries
of the five sphere. To apply the approach
presented in section (\ref{second})
we need to redefine the fermions and
pure spinor in  such a way that 
they become neutral under the isometries.
The twisted boundary conditions
for the original angles
of $AdS_5\times S^5$ will
induce twisted boundary conditions
for the original charged
fermions and pure spinors of
$AdS_5\times S^5$. 

To proceed we will
closely follow 
\cite{Alday:2005gi} since it turns
out that the approach presented
here can be easily extended to the
pure spinor string as well. 
We  begin with the study of the
 invariance
of the Lagrangian under the
abelian subalgebra of the
superconformal group. The
bosonic symmetry algebra
$SO(4,2)\times SO(6)$ has
six Cartan generators: three
for $SO(4,2)$ and three
for $SO(6)$. If we introduce
the polar representation
\begin{equation}
u_i=r_ie^{i\phi_i} \ , \quad
v_i=\rho_i e^{i\psi_i} \ \ , 
\end{equation}
where $r_i,\rho_i$ are real,
then the six commuting
isometries are realised
as constant shift of the
angle variables
\begin{equation}
\phi'_i=\phi_i+\epsilon_i \ , 
\quad  
\psi_i'=\psi_i+\tilde{\epsilon}_i \ . 
\end{equation}
It is remarkable that matrices
$g_s$ and $g_a$ enjoy
the following factorisation
property 
\cite{Frolov:2005dj,Arutyunov:2003uj,Arutyunov:2003za}
\begin{eqnarray}
g_s(r,\phi)&=&
M(\phi)\hg_s(r)M(\phi)
\ , \nonumber \\
g_a(r,\psi)&=&
M(\psi)\hg_a(\rho)
M(\psi) \ , 
\nonumber \\
\end{eqnarray}
where
\begin{equation}\label{hgsa}
\hg_s(r)=\left(\begin{array}{cccc}
0 & r_3 & r_1 & r_2 \\
-r_3 & 0 & r_2 & -r_1 \\
-r_1 & -r_2 & 0 & -r_3 \\
-r_2 & r_1 & r_3 & 0 \\
\end{array}\right) \ , \quad 
\hg_a(r)=\left(\begin{array}{cccc}
0 & \rho_3 & \rho_1 & \rho_2 \\
-\rho_3 & 0 & \rho_2 & -\rho_1 \\
-\rho_1 & -\rho_2 & 0 & \rho_3 \\
-\rho_2 & \rho_1 & -\rho_3 & 0 \\
\end{array}\right) \ , 
\end{equation}
and where $M(\phi)=e^{\frac{i}{2}\Phi}$
with $\Phi=\mathrm{diag}(\Phi_1,\dots,
\Phi_4)$ where  $\Phi_i$
 are equal to
\begin{eqnarray}
\Phi_1 & = & \phi_1+\phi_2+\phi_3 \ ,  \nonumber \\
\Phi_2 & = & -\phi_1-\phi_2+\phi_3 \ , \nonumber  \\
\Phi_3 & = & \phi_1-\phi_2-\phi_3   \ , \nonumber \\
\Phi_4 & = & -\phi_1+\phi_2-\phi_3 \ . 
 \nonumber  \\  
\end{eqnarray}
Note that in this case
the matrix $G,\tG$ can be written
as
\begin{equation}\label{Gmhg}
G=M\hG M \ , 
\quad  
 \hG=\left(\begin{array}{cc}
 \hg_a & 0 \\
 0 & \hg_s \\ \end{array}\right) \ , 
\end{equation}
\begin{equation}\label{tGmhgt}
\tG=M\htG M \ , \quad 
\htG=\left(\begin{array}{cc}
 \hg_a & 0 \\
 0 & -\hg_s \\ \end{array}\right) \ , 
\end{equation}
where
\begin{equation}\label{Mpsiphi}
M=\left(\begin{array}{cc}
M(\psi) & 0 \\
0 & M(\phi) \\ \end{array}
\right)  \ . 
\end{equation}
If we insert (\ref{Gmhg}) and
(\ref{tGmhgt}) to the action
(\ref{LprF})
we obtain that the action explicitly
depends on  $\Phi$. This fact precludes
to perform the analysis given in 
section (\ref{second}). 
In order to obtain the sigma model 
when the fermions and pure spinors
 are spectators we have
to perform  their redefinition. 

In order to find the fermionic
and pure spinor redefinition
note that fermions and pure spinor
matrices can be
written as
\begin{equation}
\theta=\left(\begin{array}{cc}
0 & X_\theta \\
Y_\theta & 0 \\ \end{array}\right) \ , 
\quad 
\lambda=\left(\begin{array}{cc}
0 & X_\lambda \\
Y_\lambda & 0 \\ \end{array}\right) \ , 
\quad 
\hlambda=\left(\begin{array}{cc}
0 & X_{\hlambda} \\
Y_{\hlambda} & 0 \\ \end{array}\right) \ , 
\end{equation}
where in the case of $\lambda,\hlambda$
the off-diagonal matrices 
$X_\lambda,Y_\lambda,X_{\hlambda},
Y_{\hlambda}$ are bosonic. We must however
stress that $\lambda,\hlambda$ are not
odd matrices of $\mathbf{su}(2,2|4)$
superalgebra. This follows from the fact that
they are defined as $\lambda=\lambda^\alpha T_\alpha$
where crucially $\lambda^\alpha$ is complex
number while for an element from $\mathbf{su}(2,2|4)$
this parameter should be real. In fact
it can be easily seen that
if $\lambda^\alpha $ were real 
the solution of the pure spinor
constraint would be trivial. 

In case of fermions  we perform
 following rescaling
\begin{equation}
X_\theta=M(\psi)\hX M(\phi)^{-1}  \ , 
\quad 
Y_\theta=M(\phi) \hX M(\psi)^{-1} \ .
\end{equation}
 Then it follows that
 \begin{equation}
 g(\theta)= M\hg(\htheta) M^{-1} \  , 
 \end{equation}
 where we have defined
 \begin{equation}
 M\equiv\left(\begin{array}{cc}
 M(\psi) & 0 \\
 0 & M(\phi) \\ \end{array}\right) \ , 
 \end{equation}
 and  where the fermions $\htheta$
are uncharged under all $U(1)$s.
Using this redefinition the currents
(\ref{J0gb2}) take the form
\begin{eqnarray}\label{J0gb2R}
\bJ^{(0)}&=&
\frac{1}{2}M(\hB-\hG \hB^T \hG^{-1}-
d\hG \hG^{-1}-
\frac{i}{2}d\Phi
-\frac{i}{2}\hG d\Phi\hG^{-1})M^{-1}\equiv
M\hbJ^{(0)}M^{-1} \  , 
\nonumber \\
\bJ^{(2)}
&=&\frac{1}{2}M(\hB+\hG \hB^T \hG^{-1}+
\frac{i}{2}d\Phi
+\frac{i}{2}\hG d\Phi\hG^{-1})M^{-1}\equiv
M\hbJ^{(2)}M^{-1}
 \ ,
\nonumber \\
\bJ^{(1)}&=&
\frac{1}{2}M[\hF-\htG \hF^T
\hG]M^{-1}\equiv M\hbJ^{(1)}M^{-1} \ ,
\nonumber \\
\bJ^{(3)}&=&
M[\hF+i\htG \hF^T \hG^{-1}]M^{-1}
\equiv M\hbJ^{(3)}M^{-1} \ 
\nonumber \\
\end{eqnarray}
and consequently the matter part
of the pure spinor action takes the form
\begin{eqnarray}\label{Lrede}
S&=&-\frac{\sqrt{\lambda}}{2\pi}\int
d\tau d\sigma\sqrt{-\eta}\str[\frac{1}{2}
\eta^{\mu\nu}\left(\hbJ_\mu^{(2)}
\hbJ_{\nu}^{(2)}+\hbJ_\mu^{(1)}\hbJ_\nu^{(3)}
+\hbJ_\mu^{(3)}\hbJ_\nu^{(1)}\right)+
\nonumber\\
&+&\frac{\epsilon^{\mu\nu}}{4}
\left(\hbJ^{(1)}_\mu \hbJ^{(3)}_\nu-
\hbJ^{(3)}_\mu \hbJ^{(1)}_\nu\right)] \ .  
\nonumber \\
\end{eqnarray}
It is important that the 
action (\ref{Lrede})
 depends on $\Phi$ through the
expression of $d\Phi$ only
and hence it is  invariant under the shift
$\Phi'=\Phi+\epsilon$. In other words
matter part of the pure spinor string
in the $AdS_5\times S^5$ background
takes the form of the sigma model
action studied in the section (\ref{second})
and consequently the TsT transformation
can be performed. 

In the similar  way as in case
of fermions we propose
the following
redefinition of the ghost
variables 
\begin{equation}\label{lambdaM}
\olambda=M\tlambda M^{-1} \ , \quad
\uhlambda=M\thlambda M^{-1} \ 
\end{equation}
and 
\begin{equation}\label{wM}
\tw_\mu= M\ow_\mu M^{-1} \ , \quad 
\thw_\mu=M \uhw_\mu M^{-1} \ ,  
\end{equation}
where $\tlambda,\thlambda,
\tw_\mu,\thw_\mu$ 
are not charged under $U(1)$'s
isometries. Note that 
(\ref{lambdaM}) and (\ref{wM})
imply that  
 $\tilde{N}_\mu,\tilde{\hN}_\mu$ 
are neutral under $U(1)'s$ isometries
as well. Further, if we insert 
(\ref{lambdaM}) into the
pure spinor constraint
we obtain 
\begin{equation}
\pb{M\tlambda M^{-1},M\tlambda M^{-1}}=
M\pb{\tlambda,\tlambda}M^{-1}=0
\end{equation}
and hence $\tlambda$  obeys the
pure spinor constraints.
It is also clear that this analysis holds
for $\tilde{\hlambda}$ as well.  
Finally we  determine the
form of $\onabla_\mu\olambda$
\begin{eqnarray}
\onabla_\mu\olambda&=&
\partial_\mu\olambda+\com{\bJ^{(0)}_\mu,\olambda}\nonumber \\
&=&M(\partial_\mu\tlambda+\frac{i}{2}
[\partial_\mu\Phi,\tlambda]+
[\hbJ^{(0)}_\mu,\tlambda])M^{-1}\equiv
M\hnabla_\mu \tlambda M^{-1}
\nonumber \\
\end{eqnarray}
using
\begin{eqnarray}
d\olambda= 
M\left (d\tlambda+ \frac{i}{2}d\Phi \tlambda 
-\frac{i}{2}\tlambda d\Phi\right)M^{-1} \ . 
\nonumber \\
\end{eqnarray}
Clearly the same equation  holds for 
$\uhlambda,\uhw_\mu $. In summary we obtain
following form of the pure spinor Lagrangian
from (\ref{LprF})
\begin{eqnarray}\label{tghost}
\mL_{pure}=-\frac{\sqrt{\lambda}}{2}
\str[\tw_\mu \cP^{\mu\nu}
\hnabla_\nu \tlambda+
\thw_\mu \tP^{\mu\nu}\hnabla_\nu \thlambda
-\tilde{N}_\mu \tP^{\mu\nu}
\tilde{\hat{N}}_\nu] \ . 
\nonumber \\
\end{eqnarray}
We see that (\ref{tghost})
 depends on 
$\Phi$ through 
$d\Phi$ only and 
hence the analysis performed in
section (\ref{second}) can be applied
for pure spinor action as well. 

It will be useful to express the
equations of motion for $\bJ$ and ghosts 
$\olambda,\uhlambda,\ow,\uhw$ 
that were  given in
(\ref{eqm1r}),(\ref{eqm2r}),
(\ref{eqm3ar}),(\ref{eqm3br}),
(\ref{eqg1r}) and (\ref{eqg2r})
in terms of the variables
defined in (\ref{J0gb2R}),(\ref{lambdaM})
and (\ref{wM}).
As the first step we express the
covariant derivative $\onabla \bJ^{(i)}$
using the redefined currents
(\ref{J0gb2R})
\begin{eqnarray}\label{onabjaN}
\onabla \bJ^{(i)}
=M(d\hbJ^{(i)}+
\frac{i}{2}\com{d\Phi,\hbJ^{(i)}}+
\com{\hbJ^{(0)},
\hbJ^{(i)}})M^{-1}\equiv
M\hnabla \hbJ^{(i)}M^{-1} \ ,
\nonumber \\
\end{eqnarray}
where we have introduced the
derivative $\hnabla$ that by definition
depends on $\hbJ^{(0)}$ and 
on the derivative of $\Phi$. 
Then with the help of
(\ref{J0gb2R}),(\ref{lambdaM}),(\ref{wM})
and (\ref{onabjaN}) we can
determine from  (\ref{eqm1r}),(\ref{eqm2r}),
(\ref{eqm3ar}),(\ref{eqm3br}),
(\ref{eqg1r}) and (\ref{eqg2r})
the equations of motion 
for $\hbJ,\tlambda,\thlambda,\tw_\mu$
and $\thw_\mu$ in the form
\begin{eqnarray}\label{eqR3}
\tP^{\mu\nu}
\hnabla_\mu \hbJ^{(3)}_\nu
+
[\hbJ_\nu^{(3)},\tilde{N}_\mu]\cP^{\mu\nu}+
[\hbJ_\nu^{(3)},\tilde{\hN}_\mu]\tP^{\mu\nu}
=0 \ ,
\nonumber \\
\cP^{\mu\nu}\hnabla_\mu \hbJ^{(1)}_\nu
+
[\hbJ_\nu^{(1)},\tilde{N}_\mu]\cP^{\mu\nu}+
[\hbJ_\nu^{(1)},\tilde{\hN}_\mu]\tP^{\mu\nu}
=0 \ , \nonumber \\
\cP^{\mu\nu}
\hnabla_\mu \hbJ^{(2)}_\nu
-\epsilon^{\mu\nu}
[\hbJ^{(1)}_\mu,\hbJ^{(1)}_\nu]
+[\hbJ_\nu^{(2)},\tilde{N}_\mu]\cP^{\mu\nu}+
[\hbJ_\nu^{(2)},\tilde{\hN}_\mu]\tP^{\mu\nu}
=0 \ , \nonumber \\
\tP^{\mu\nu}
\hnabla_\mu \hbJ^{(2)}_\nu
+\epsilon^{\mu\nu}
[\hbJ^{(3)}_\mu,\hbJ^{(3)}_\nu]
+[\hbJ_\nu^{(2)},\tilde{N}_\mu]\cP^{\mu\nu}+
[\hbJ_\nu^{(2)},\tilde{\hN}_\mu]\tP^{\mu\nu}
=0 \ ,\nonumber \\
\cP^{\mu\nu}\hnabla_\nu\tlambda+
\cP^{\mu\nu}[\tlambda,\tilde{\hN}_\nu]
=0 \ , \nonumber \\
\tP^{\mu\nu}\hnabla_\nu \thlambda
+\tP^{\mu\nu}[\thlambda,\tilde{N}_\nu]
=0 \ . \nonumber \\
\end{eqnarray}
Finally we will discuss the
conservation of the  
BRST currents
given in (\ref{jrl}).
 With the help of 
(\ref{J0gb2R}) and (\ref{lambdaM})
 it is easy
to see that they are  equal to 
\begin{equation}\label{jrlM}
\tj^\mu_R=\str(\thlambda \hbJ_\nu^{(1)}
\tP^{\nu\mu}) \ , \quad
\tj^\mu_L=\str(\tlambda \hbJ_\nu^{(3)}
\cP^{\nu\mu}) \  
\end{equation}
and that they are again conserved as
a consequence of the equations
of motion (\ref{eqR3}). 
Consequently the  time
derivative of the BRST charges
is equal to
\begin{equation}\label{dQtt}
\frac{dQ_L}{d\tau}=-\frac{1}{2\pi}
(\tj^1_L(2\pi)-\tj^1_L(0)) \ , \quad 
\frac{dQ_R}{d\tau}=-\frac{1}{2\pi}
(\tj^1_R(2\pi)-\tj^1_R(0)) \ .
\end{equation}
\section{TsT transformation
on the five sphere}\label{sixth}
Even if the general analysis performed
above is valid for the TsT transformation
in the $AdS_5$ space as well
we restrict  to the
 TsT transformation
applied to
the five-sphere, 
following \cite{Alday:2005gi}.  
 This restriction
implies that we do need to impose
that fermions and pure spinors
are neutral under isometries of
$AdS_5$. Then we can take $M(\psi)=1$
and hence we obtain that the matrix
$M$ takes the form
\begin{equation}
M=\left(\begin{array}{cc}
1 & 0 \\
0 & M(\phi) \\ \end{array}\right)
 \ . 
\end{equation}

In order to determine the
twisted boundary conditions
for fermions and pure spinors
we have to take into account
that the redefined fermions and
the pure spinors do not transform
under the TsT transformations.
Therefore the original charged
fermions in $AdS_5\times S^5$
and pure spinors satisfy the
twisted boundary conditions. We
find these boundary conditions
using the relation between
$\htheta$ and $\theta$ and
the twisted boundary condition for angle
$\phi_i$ that has impact
on the matrix $M$
since 
\begin{equation}
\phi_i(2\pi)=\phi_i(0)+2\pi (n_i-\nu_i) \ , \quad 
\nu_i=\epsilon_{ijk} \gamma_j J_k
\end{equation}
or equivalently 
\begin{eqnarray}\label{Phi2pi}
\Phi_1(2\pi)=\Phi_1(0)+2\pi(n_1+n_2+n_3-
\nu_1-\nu_2-\nu_3)\equiv
\Phi_1(0)-2\pi\Lambda_1 \ , \nonumber \\
\Phi_2(2\pi)=\Phi_2(0)+2\pi(-n_1-n_2+n_3
+\nu_1+\nu_2-\nu_3)\equiv 
\Phi_2(0)-2\pi\Lambda_2 \ , \nonumber \\
\Phi_3(2\pi)=\Phi_3(0)
+2\pi(n_1-n_2-n_3-\nu_1+\nu_2+\nu_3)
\equiv \Phi_3(0)-2\pi\Lambda_3 \ ,
\nonumber \\
\Phi_4(2\pi)=\Phi_4(0)
+2\pi(-n_1+n_2-n_3+\nu_1-\nu_2+\nu_3)
\equiv \Phi_4(0)-2\pi\Lambda_4 \ .
\nonumber \\
\end{eqnarray}
Using (\ref{Phi2pi}) we easily
obtain 
\begin{eqnarray}
M(\Phi(2\pi))=\left(\begin{array}{cccc}
e^{-i\pi\Lambda_1} & 0 & 0 & 0 \\
0 & e^{-i\pi\Lambda_2} & 0 & 0  \\
 0 & 0 & e^{-i\pi\Lambda_3} & 0\\
0 & 0 & 0 & e^{-i\pi\Lambda_4} \\
\end{array}\right)
\left(\begin{array}{cccc}
e^{i\Phi_1(0)} & 0 & 0 & 0 \\
0 & e^{i\Phi_2(0)} & 0 & 0  \\
 0 & 0 & e^{i\Phi_3(0)} & 0\\
0 & 0 & 0 & e^{i\Phi_4(0)} \\
\end{array}\right) 
\nonumber \\
\end{eqnarray}  
or in compact notation
\begin{equation}\label{MPHi2pi}
M(\Phi(2\pi))=e^{-i\pi\Lambda}M(\Phi(0))
\ , 
\end{equation}
where $\Lambda=\mathrm{diag}(\Lambda_1,\Lambda_2,
\Lambda_3,\Lambda_4)$. 
Then we have
\begin{eqnarray}
g(\theta)(2\pi)&=&
M(2\pi)g(\htheta)(2\pi)M^{-1}(2\pi)\nonumber \\
&=& \left(\begin{array}{cc}
 1 & 0 \\
 0 & e^{-i\pi \Lambda}
 \\ \end{array}\right)
 M(\Phi(0))
 g(\htheta(0))M^{-1}(\Phi(0))
 \left(\begin{array}{cc}
 1 & 0 \\
 0 & e^{i\pi \Lambda}
 \\ \end{array}\right)
\nonumber \\
&=&\left(\begin{array}{cc}
1 & 0 \\
0 & e^{-i\pi \Lambda}
\\ \end{array}\right)
g(\theta)(0)
\left(\begin{array}{cc}
1 & 0 \\
0 & e^{i\pi \Lambda}
\\ \end{array}\right)
\nonumber \\
\end{eqnarray}
using the fact that $\htheta$
do not transform under TsT duality and
hence they are the same in TsT dual background
with standard periodicity $\htheta(2\pi)=\htheta(0)$. 

Now we would like to  
explain  carefully our calculations. 
We have derived
 the pure
spinor action in the $AdS_5\times S^5$ 
given in (\ref{Lrede}) and (\ref{tghost}) that
by construction is
  manifestly invariant
under the isometry of the background parametrised
by $\Phi$.  Now let
us suppose that we have pure
spinor action that describes
\emph{closed string}  in the
$\gamma$-deformed background. 
Since the $\gamma$-deformed background
can be derived from the original $AdS_5\times
S^5$ by sequence of the TsT transformations
the analysis performed in section 
(\ref{second}) suggests that this
action has the same  functional form as
the action given  in (\ref{Lrede}) and (\ref{tghost}).
Let us  denote the corresponding Lagrangian
as  $\mL(\hbJ^0,\tlambda,
\thlambda)$ where superscripts on
$\hbJ$ mean that these currents explicitly
depend on the $\gamma$-deformed background. 
According to the 
arguments given in the section (\ref{second})
this Lagrangian can be mapped by sequence of TsT
transformations to the Lagrangian 
$\mL(\hbJ,\tlambda,
\thlambda)$ where now the angle variables
obey twisted boundary conditions
according to (\ref{MPHi2pi}). On the other
hand the fermionic 
 $\htheta$ and ghost variables
$\tlambda,\thlambda,\tw,\thw$ have periodic
boundary conditions since they are
 neutral under $U(1)'s$ isometries.
Then the form of the currents
$\hbJ$  given
in (\ref{J0gb2R}) imply that they are periodic
since they depend on $r,\rho$ and
$\htheta$ and as we argued
above these modes are periodic. 
 It is also easy
to see that $d\Phi(2\pi)=d\Phi(0)$. 
Explicitly, we have 
\begin{eqnarray}
\hbJ^{(i)}(2\pi)&=&\hbJ^{(i)}(0) \ , \quad 
i=0,1,2,3 \ , \nonumber \\
\tlambda(2\pi)&=&\tlambda(0) \ , \quad
\thlambda(2\pi)= \thlambda(0) \ , \quad
\nonumber \\
\tw_\mu(2\pi)&=&\tw_\mu(0)\ , \quad 
\thw_\mu(2\pi)=\thw_\mu(0) \ . \nonumber \\
\end{eqnarray}
These boundary conditions immediately show that
the conserved  BRST currents given 
in (\ref{jrlM}) imply the existence of
the time-independent BRST charges as
follows from (\ref{dQtt}).  

On the other hand we can take one step further
and study the pure spinor action expressed
with the variables $\bJ,\olambda,\uhlambda$. 
These variables now obey twisted boundary conditions
as follows from (\ref{J0gb2R}),(\ref{lambdaM})
and (\ref{wM})
\begin{eqnarray}\label{tWbJ}
\bJ^{(i)}(2\pi)&=&
\left(\begin{array}{cc}
1 & 0 \\
0 & e^{-i\pi \Lambda}
\\ \end{array}\right)
\bJ^{(i)}(0)
\left(\begin{array}{cc}
1 & 0 \\
0 & e^{i\pi \Lambda}
\\ \end{array}\right) \ ,
\quad 
 i=0,1,2,3
\nonumber \\ 
\olambda(2\pi)&=&
\left(\begin{array}{cc}
1 & 0 \\
0 & e^{-i\pi \Lambda}
\\ \end{array}\right)
\olambda(0)
\left(\begin{array}{cc}
1 & 0 \\
0 & e^{i\pi \Lambda}
\\ \end{array}\right) \ , \quad 
\uhlambda(2\pi)=\left(\begin{array}{cc}
1 & 0 \\
0 & e^{-i\pi \Lambda}
\\ \end{array}\right)
\uhlambda(0)
\left(\begin{array}{cc}
1 & 0 \\
0 & e^{i\pi \Lambda}
\\ \end{array}\right) \ , 
\nonumber \\
\ow_\mu(2\pi)&=&
\left(\begin{array}{cc}
1 & 0 \\
0 & e^{-i\pi \Lambda}
\\ \end{array}\right)
\ow_\mu(0)
\left(\begin{array}{cc}
1 & 0 \\
0 & e^{i\pi \Lambda}
\\ \end{array}\right) \ , \quad 
\uhw_\mu(2\pi)=\left(\begin{array}{cc}
1 & 0 \\
0 & e^{-i\pi \Lambda}
\\ \end{array}\right)
\uhw_\mu(0)
\left(\begin{array}{cc}
1 & 0 \\
0 & e^{i\pi \Lambda}
\\ \end{array}\right) \ . 
\nonumber \\
\end{eqnarray}
We again see that these boundary
conditions  immediately
show that the conserved BRST
currents (\ref{jrl}) are periodic
and hence they are two 
time-independent BRST charges
as follows from
(\ref{dtQ}). In other words we
have shown that classically the pure
spinor string is well defined
even in the case when the world-volume
fields obey the twisted boundary
conditions. 
On one hand  
the power of pure spinor formalism is
that it allows to prove exact conformal
invariance of the pure spinor string in
$AdS_5\times S^5$ background
\cite{Berkovits:2004xu} and 
arguments given there crucially depend
on the gauge invariance of the pure
spinor string with respect to subgroup
$SO(4,1)\times SO(5)$. On the other hand
the form of the action (\ref{LprF}) 
explicitly depends on the coset representative
that corresponds to the fixing of
the gauge $SO(4,1)\times SO(5)$.
Consequently the
action (\ref{LprF})  is not suitable for the
analysis of the general properties of the
pure spinor sigma model with twisted boundary
conditions.  

To find such a formulation we
would like to express the theory where
the fundamental fields obey the
twisted boundary conditions in terms
of the original currents $J$ and ghost
variables that 
appear in the action (\ref{Minaction})
and that  obey 
some form of the twisted boundary 
conditions:
\begin{eqnarray}\label{boundh}
J(2\pi)&=&N(\Lambda)J(0)N^{-1}(\Lambda) \ , 
\nonumber \\
\lambda(2\pi)&=&N(\Lambda)\lambda(0)N^{-1}(\Lambda) \ , 
\quad 
\hlambda(2\pi)=N(\Lambda)\hlambda(0)N^{-1}(\Lambda) \ ,
\nonumber \\
w_\mu(2\pi)&=&N(\Lambda)w_\mu(0) N^{-1}(\Lambda) \ , 
\quad \hw_\mu(2\pi)= N(\Lambda)\hw_\mu(0)
N^{-1}(\Lambda) \ , \nonumber \\
\end{eqnarray}
for some matrix $N$ that depends on 
$\Lambda$ only. If we were able to
find such a formulation then we would
get the original action (\ref{Minaction})
with the explicit gauge invariance but
where  now the world-volume fields 
obey the twisted  boundary conditions 
(\ref{boundh}). Since the 
algebraic renormalisation 
arguments given \cite{Berkovits:2004xu} (see also 
\cite{piguet}
are sensitive to the UV properties
of the theory we could then argue
that (\ref{Minaction}) with 
fields obeying the twisted 
boundary conditions 
(\ref{boundh}) defines exact
quantum field theory. It turns out however
that it is not possible to find
such a form of the boundary conditions.

To be more precise we try to 
find the boundary conditions of the original
currents $J^{(i)} \ , i=1,2,3$ and
ghosts $\lambda,\hlambda$ using the
relations (\ref{lambdare}) and
(\ref{J0gb2}). These relations imply
that $J,\lambda,\hlambda$ explicitly
depend on $g$ that has the form
\begin{equation}
g=\left(\begin{array}{cc}
\tilde{g}_a & 0 \\
0 & \tilde{g}_s \\ 
\end{array}\right)
\end{equation}
where $\tilde{g}_a(2\pi)=\tilde{g}(0)$ as follows
from the fact that $g_a$ parametrises
$AdS_5$. 
More difficult problem is to find the
boundary condition for $g_s$. Recall that
this group element has the form
\begin{equation}
\tilde{g}_s=\exp\left(\frac{i}{2}y_a \Gamma_a\right) \ ,
\end{equation} 
where $y_a$ parametrise the five-sphere and
$\Gamma_a, a=1,\dots,5$ are the Dirac matrices
for $SO(5)$. The variables $y_a$
are related to $\rho,\phi$ given in 
(\ref{hgsa}) as
\begin{eqnarray}\label{cosu}
y_1&=&\frac{1}{2}
\frac{|y|}{\sin|y|}(u_1-u_1^*) \ , \quad
y_2=-\frac{1}{2}
\frac{|y|}{\sin|y|}(u_2+u_2^*) \ , 
\nonumber \\
y_3&=&\frac{1}{2}
\frac{|y|}{\sin|y|}(u_2-u_2^*) \ , \quad  
y_4=-\frac{1}{2}
\frac{|y|}{\sin|y|}(u_1+u_1^*) \ , 
\nonumber \\
y_5&=&\frac{1}{2}
\frac{|y|}{\sin|y|}(u_3-u_3^*) \ ,  \quad 
|y|=-\sin^{-1}\left(
 \frac{u_3+u^*_3}{2}\right) \ , \nonumber \\
\quad |y|^2&=&
y_1^2+y_2^2+y_3^2+y_4^2+y_5^2 \ .  
\nonumber \\
\end{eqnarray}
Then using the fact that
$g_s$ obeys following
boundary conditions
\begin{equation}
g_s(2\pi)=e^{-i\pi\Lambda}g_s(0)e^{-i\pi\Lambda}
\end{equation}
we easily find the twisted
boundary conditions for
 $u_i$ 
\begin{eqnarray}
u_1(2\pi)&=&e^{-i\pi(\Lambda_1+\Lambda_3)}u_1(0) \ , 
\nonumber \\
u_2(2\pi)&=&e^{-i\pi(\Lambda_1+\Lambda_4)}u_2(0) \ , 
\nonumber \\
u_3(2\pi)&=&e^{-i\pi(\Lambda_1+\Lambda_2)}u_3(0) \ . 
\nonumber \\
\end{eqnarray}
Using the boundary conditions for
$u_i$ given above we can 
easily  determine the boundary condition
for $|y|$ given in (\ref{cosu})
\begin{equation}
|y|(2\pi)=\sin^{-1}\left[2
\sin |y|(0)\cos(\pi(\Lambda_1+\Lambda_2)+
2i\sin (\pi(\Lambda_1+\Lambda_2)
\frac{y_5(0)}{|y|(0)}\right] \ . 
\end{equation}
This result clearly demonstrates
 that it is not
possible to find the appropriate $N$ matrix
introduced in (\ref{boundh}) for general
values of $\Lambda$.  This is a consequence
of the fact that the global symmetries
of the coset are realised non-linearly on
coordinates $y$. 

Since we have shown that is not possible
find the matrix $N$ for general $\Lambda$
it turns out that 
the arguments in \cite{Berkovits:2004xu}
 that were based
on the existence of local gauge symmetry
cannot be applied to the currents that
obey twisted boundary conditions.  
However, looking at the above equations we see
that for $\nu_i\in  Z$, where
$\nu_i$ is defined as  
\begin{equation}\label{nujk}
\nu_i=\epsilon_{ijk}\gamma_j J_k \  
\end{equation}
the world-sheet fields in the
original  $AdS_5\times S^5$
background  obey the periodic boundary 
conditions as well. In fact
this requirement is in agreement with the
analysis performed in 
 \cite{Frolov:2005iq} where the importance
 of the solution with integer $\nu_i$ 
 was stressed. 
We are not going to perform the same analysis
since we have not studied the classical solutions
of the pure spinor string in the 
original 
$AdS_5\times S^5$ background however we would
like to stress some interesting points considering
the condition that $\nu_i$ is an integer. 
For $\nu_i\neq 0$ we  have consistent
string dynamics if $\gamma_i$ are rational since
$J_i$ take integer values in quantum theory. 
Secondly, the condition $\nu_i=0$
has the general solution
\begin{equation}
\nu_i=0: \quad J_i=c\gamma_i \ . 
\end{equation}
Since again $J_i$ have to be integer
in quantum theory these solutions exist
for special values of $\gamma_i$. 

Returning back to (\ref{cosu}) we
see that the matrix $g_s$ is periodic. 
Then using also the fact that $\bJ^{(i)}$
are periodic as well we obtain that
the currents $J$ given in (\ref{Minaction})
obey the standard boundary conditions.
In other words we can formulate the dynamics
of the theory in terms of the original currents
$J$ and the action (\ref{Minaction}) is manifestly
gauge invariant.  According
to the analysis given in  
\cite{Berkovits:2004xu} the pure spinor
string action in 
$AdS_5\times S^5$ possesses
 quantum BRST invariance and
also exact conformal invariance. 
Then using the TsT transformations
we can map the configurations of the
pure spinor string in $AdS_5\times S^5$
to the  states in the $\gamma_i$-deformed background
that obey the condition 
$\nu_i$ is an integer 
and we can expect that these states
are exact states even in the quantum theory 
of the pure spinor string in the $\gamma$-deformed
background. 

Since we found that the proof of the conformal invariance  
of \cite{Berkovits:2004xu} strictly depends on the manifest isometries 
of the background, in other -- more realistic -- situations 
(for instance, N=1 supersymmetries backgrounds) 
another way of proving the conformal invariance has to be developed. 

\section{Lax Pair for twisted pure spinor string}
\label{seventh}
Our goal is to find, using the relations
(\ref{relphitphi}) the Lax pair for string
in TsT transformed background if an
isometry invariant Lax pair for pure spinor
string in flat background is known. 
An  existence
of Lax pair in deformed theory  
 strongly supports
classical integrability of the theory
\cite{Berkovits:2004jw,Vallilo:2003nx,Bena:2003wd,
Alday:2003zb,Alday:2005gi,
Das:2004hy,Kazakov:2004qf,Arutyunov:2004yx}.

We begin with the recalling the structure
of Lax pair for pure spinor string in
$AdS_5\times S^5$. 
In the covariant pure spinor formalism 
the problem has been
studied in  \cite{Vallilo:2003nx}. 
It was shown here that  there
exists 
set of left-invariant currents
 $\hat{J}(u)$ 
\footnote{Our spectral parameter $u$
is related to the spectral parameter $\mu$ of
\cite{Vallilo:2003nx} by $\mu= e^u$.
Note also that we have chosen one particular solution from
the ones found in \cite{Vallilo:2003nx} in order to 
obey the initial condition $\hat{J}_\mu(0)=J_\mu$. 
It is remarkable that the classical theory admits the same two
one-parameter families of
flat currents if one sets the contribution of the pure spinor
ghost $N$ to zero.} 
\begin{eqnarray}\label{flato}
 \hat{J}_\mu(u) &=&J_\mu+ ( \eta_{\mu\nu} (\cosh{u} -1)+
\epsilon_{\mu\nu} \sinh{u}) J^{\nu(2)}+
\nonumber \\
&+&( \eta_{\mu\nu} (\cosh{u}e^{u/2} -1)+
\epsilon_{\mu\nu} \sinh{u}e^{u/2}) J^{\nu(1)}
+\nonumber \\
&+&( \eta_{\mu\nu} (\cosh{u} e^{-u/2}-1)+
\epsilon_{\mu\nu} \sinh{u} e^{-u/2}) J^{\nu(3)}
 +\nonumber \\
&+&\sinh u e^{u}\tP_{\mu\nu} N^{\nu}
-\sinh ue^{-u}\cP_{\mu\nu}\hN^{\nu} \  
\nonumber \\ 
\end{eqnarray}
that satisfy   the
flatness condition 
\begin{equation}\label{flatnes}
 d\hat{J} + \hat{J} \wedge \hat{J} = 0 
\end{equation}
that is a consequence of the
equations of motion for $J$ and 
ghost fields and also of the flatness
of $J$. Note also that $\hJ$
obeys the   the 'initial'
condition $\hat{J}(0)= J$.

The Lax connection given above cannot
be used to derive the Lax connection
in deformed background since $J_\mu^{(i)}$
given there explicitly depend on $\phi$ and
consequently $\hJ_\mu$ is not
isometry invariant.  Moreover, if we
express  $J^{(0)}$  using (\ref{J0gb})
it turns out that it 
explicitly  depends on $g(z)$ and
it is not clear how to related the original
Lax connection in the $AdS_5\times S^5$
background to the Lax connection in TsT transformed
one. To resolve this problem we will
proceed in the similar way as in    
 \cite{Frolov:2005dj,Alday:2005ww,
Alday:2005gi,Beisert:2005bm}.
Let us write the flat current $\hJ$ as
\begin{equation}
\hJ=g^{-1}(z)dg(z)+g^{-1}(z)\hJ' g(z) \ . 
\end{equation}
Then
the flatness of $\hJ$ implies  
\begin{eqnarray}
d\hJ+\hJ\wedge \hJ
=g^{-1}(d\hJ'+\hJ'\wedge \hJ')g=0 \  
\nonumber \\
\end{eqnarray}  
and hence $d\hJ'+\hJ'\wedge \hJ'=0$. Now
(\ref{flato}) implies
\begin{eqnarray}\label{hjtr}
 \hJ'_\mu(u)&=&g(z)\hJ_\mu g^{-1}(z)-
 \partial_\mu g(z)g^{-1}(z) \nonumber \\
&=&\bJ_\mu+[ ( \eta_{\mu\rho} (\cosh{u} -1)+
\epsilon_{\mu\rho} \sinh{u})\eta^{\rho\sigma}\bJ_\sigma^{(2)}
+
\nonumber \\
&+&( \eta_{\mu\rho} (\cosh{u}e^{u/2} -1)+
\epsilon_{\mu\rho} \sinh{u}e^{u/2}) 
\eta^{\rho\sigma}\bJ^{(1)}_\sigma
+\nonumber \\
&+&( \eta_{\mu\rho} (\cosh{u} e^{-u/2}-1)+
\epsilon_{\mu\rho} \sinh{u} e^{-u/2}) 
\eta^{\rho\sigma}\bJ^{(3)}_\sigma
 +\nonumber \\
&+&\sinh u e^{u}\tP_{\mu\nu} \overline{N}^{\nu}
-\sinh ue^{-u}\cP_{\mu\nu}\underline{\hN}^{\nu} ] \ , 
\nonumber \\
\end{eqnarray}
where we have also used 
(\ref{lambdare}) and
(\ref{J0gb2}). The Lax connection
$\hJ'$  (\ref{hjtr}) still has explicit
dependence on $\phi^i$ but this can
be easily eliminated using the
factorisation property of $G$ and 
redefinition of the fermions and pure
spinors.
Explicitly, using  the relations 
(\ref{J0gb2R}), (\ref{lambdaM}) and
(\ref{wM})) we can write 
(\ref{hjtr}) as
\begin{eqnarray}\label{hJM}
\hJ'_\mu&=&M\hbJ_\mu M^{-1}+
M[ ( \eta_{\mu\rho} (\cosh{u} -1)+
\epsilon_{\mu\rho} \sinh{u})\eta^{\rho\sigma}\hbJ_\sigma^{(2)}
+
\nonumber \\
&+&( \eta_{\mu\rho} (\cosh{u}e^{u/2} -1)+
\epsilon_{\mu\rho} \sinh{u}e^{u/2}) 
\eta^{\rho\sigma}\hbJ^{(1)}_\sigma
+\nonumber \\
&+&( \eta_{\mu\rho} (\cosh{u} e^{-u/2}-1)+
\epsilon_{\mu\rho} \sinh{u} e^{-u/2}) 
\eta^{\rho\sigma}\hbJ^{(3)}_\sigma
 +\nonumber \\
&+&\sinh u e^{u}\tP_{\mu\nu} \tilde{N}^{\nu}
-\sinh ue^{-u}\cP_{\mu\nu}\tilde{\hN}^{\nu} ]M^{-1}
\equiv M (\hat{\bJ}_\mu-\frac{i}{2}
\partial_\mu \Phi) M^{-1} \ . 
\nonumber \\
\end{eqnarray}
Then the flatness condition for $\hJ'$ implies
\begin{eqnarray}
d\hJ'+\hJ'\wedge \hJ'
=M(d\hat{\bJ}+
\hat{\bJ}\wedge \hat{\bJ})M^{-1}=0
\nonumber \\
\end{eqnarray}
and consequently we obtain the
flatness condition for $\hat{\bJ}$
\begin{equation}
d\hat{\bJ}+
\hat{\bJ}\wedge \hat{\bJ}=0 \ .
\end{equation}
We see, following the arguments
given in section (\ref{fifth}) that
the Lax connection $\hat{\bJ}$ 
depends  on the derivatives of
$\Phi$ only. Then following the
arguments given in  
\cite{Frolov:2005dj} 
we  can determine the Lax connection for
pure spinor strings in the $\gamma$-deformed
$AdS_5\times S^5$ when we express
$\partial_\mu \phi_i$ in terms of
$\partial_\mu \tphi_i$ with the help
of the  
relations (\ref{relphitphi}) and also  using the
fact that 
$\hat{\bJ}$ depends on variables
that are neutral under $U(1)$ only.
By construction the Lax connection
$\hat{\bJ}$ is flat, it is invariant
under $U(1)$ isometries and it 
also obeys the periodic boundary conditions.
 It can be used to compute
the monodromy matrix $T(u)$ that is
defined as the path-ordered 
exponential of the spatial component of
$\hat{\bJ}_\sigma$  
\begin{equation}
T(u)=P\exp \int_0^{2\pi}
d\sigma \hat{\bJ}_\sigma (u) \ . 
\end{equation}
On the other hand we have argued that
in order to study the quantum properties
of the string theory in TsT-deformed
background it is necessary that the
world-sheet modes in the original
$AdS_5\times S^5$  
background obey the periodic boundary
conditions. This results also implies
that $\hJ'$ and $\hJ$ are periodic 
as well and their analysis can be performed
as in \cite{Berkovits:2004xu,Berkovits:2004jw}.

\section*{Acknowledgements}

J.K. would like to thank 
DISTA, Universit\`a del Piemonte Orientale   for hospitality
where  this work was initiated. 
This work  was supported in part by the Czech Ministry of
Education under Contract No. MSM
0021622409, by INFN, by the MIUR-COFIN
contract 2003-023852 and 2005-023102, by the EU
contracts MRTN-CT-2004-503369 and
MRTN-CT-2004-512194, by the INTAS
contract 03-516346 and by the NATO
grant PST.CLG.978785.


\end{document}